\documentclass{aa}
\usepackage{graphicx}

\begin{document}

\newcommand{\fe}{\ion{Fe}{ii}}
\newcommand{\h}{H$_2$}
\newcommand{\ci}{\ion{C}{i}}
\newcommand{\s}{\ion{S}{ii}}
\newcommand{\n}{\ion{N}{i}}
\newcommand{\nii}{\ion{N}{ii}}

\title{On the excitation of the infrared knots along protostellar jets
\thanks{Based on observations collected at the European Southern
Observatory, La Silla, Chile (66.C-158, 68.C-0020, 69.C-0175)}}
\author{T. Giannini$^1$, C. M$^{\rm c}$Coey$^2$, A. Caratti o
Garatti$^{1,3}$, B. Nisini$^1$, D. Lorenzetti$^1$, D.R. Flower$^2$}
\offprints{Teresa Giannini, email:teresa@coma.mporzio.astro.it}
\institute{$^1$ INAF-Osservatorio Astronomico di Roma, via Frascati 33,
I-00040 Monteporzio Catone,
Italy\\
$^2$ Physics Department, The University, Durham DH1 3LE, UK\\
$^3$ Universit\`a degli Studi `Tor Vergata', via della Ricerca Scientifica
1, I-00133 Roma, Italy}
%
%
\date{Received date; Accepted date}
%
%
%
\titlerunning{On the excitation of the infrared knots along protostellar jets}
\authorrunning{T.Giannini et al.}

\abstract{ The complete near infrared (0.9-2.5 $\mu$m) spectra in three
different star forming regions (HH24-26, HH72 and BHR71) are presented and
analyzed in the framework of shock excitation models. The spectra are
dominated by H$_2$ rovibrational emission (vibrational state $\nu$ $\le$
5, excitation energy $T_{ex}$ $\le$ 35\,000 K), while emission from ionized
material, recognizable from [Fe II] and [S II] lines, is significantly
fainter. The analysis of the H$_2$ excitation diagrams points to the
existence of two different excitation regimes: whilst condensations
observed only in the infrared appear to have temperatures rarely exceeding
3000 K and can be modelled in the framework of steady-state C-shock
models, the infrared counterparts of Herbig Haro (HH) objects
exhibit a temperature stratification with components up to more than 5000
K. The H$_2$ emission from representative HH objects (HH26A, HH72A
and HH320A) has been successfully modelled by planar J-shocks with
magnetic precursors, for which the main parameters (pre-shock density,
speed) are derived. However, these same models are unable to reproduce the
observed atomic and ionic emission, which probably arises from a
distinct and perhaps more embedded region with respect to that traced by
the H$_2$. Some of the physical parameters of such regions (fractional
ionization, density) have been estimated in HH72, on the basis of the
observed ionic lines.
} \maketitle 


\section{Introduction}

The physical effect exerted on the ambient medium by the impact of jets
from accreting protostars results in the formation of shock waves. The
compressed and heated gas radiates away the accumulated thermal energy
through the emission of ionic, atomic and molecular lines, in relative
proportions which depend on the structure of the shock wave. During the
gas cooling, a fundamental role is played by molecular hydrogen, whose
high abundance compensates the low transition rates due to its homonuclear
nature. In the near infrared, the association of H$_2$ emission with knots
of shocked gas has been demonstrated by mapping at sub-arcsecond scale in
the $\nu$=1-0 S(1) line at 2.122 $\mu$m (e.g. Eisl\"{o}ffel 2000), and the
rovibrational lines are effective probes of the molecular gas at thousands
of kelvin, which is missed by optical observations. H$_2$ line
intensities and intensity ratios are also extensively used to distinguish
between fluorescence and shock excitation as possible mechanisms at the
origin of the emission, since strong H$_2$ lines with $\nu$$\ge$6
in the 1.0-1.4 $\mu$m range are expected if fluorescence is responsible for 
the emission (Black $\&$ van Dishoeck 1987). It has proved more 
difficult to derive the nature (C-ontinuous or J-ump type, e.g. Kaufman $\&$ Neufeld 1996,
Hollenbach $\&$ McKee 1989, Draine 1980) and the physical parameters of
the shock waves, such as the velocity, the preshock density, the strength
of the transverse magnetic field and the temperature of the neutral gas.
In the conventional scenario, C-type shocks can exist up to velocities
typically $\approx$~50 km s$^{-1}$, beyond which the H$_2$ molecule is
collisionally dissociated and the kinetic temperature of the gas rapidly
increases, giving rise to a discontinuity in the shock parameters.
However, more recent models (e.g. Smith 1995, Le Bourlot et al. 2002,
Flower et al. 2003) have shown that H$_2$ dissociation can be inhibited
over a wider range of shock parameters: from slow ($v_{s}$ $\la$ 25 km
s$^{-1}$), partially dissociative J-type shocks, up to fast C-type shocks
travelling at 70-80 km s$^{-1}$.  From an observational point of view, the
presence of different shock components (J- or C-type) can be
demonstrated by the modifications induced in the H$_2$ excitation diagram:
the lower post-shock densities attained in C-shocks result in a larger
departure from LTE conditions than in J-shocks of the same speed (Flower
et al. 2003). In order to probe such deviations, it is essential to
investigate spectroscopically the 1.0-1.4 $\mu$m range, since several
H$_2$ lines with different vibrational quantum number and of relatively
high excitation energy ($>$ 15\,000 K) lie at these wavelengths.  With the
aim of observing such lines, we have undertaken a spectroscopic survey
(from 0.9 to 2.5 $\mu$m) of a sample of Herbig-Haro (HH) objects and H$_2$
jets, based on the observations gathered with the SOFI spectrometer at the
ESO-NTT. We have already reported the results of this survey regarding
HH43, HH111, HH240/241 and HH120 (Giannini et al. 2002, Nisini et al.
2002, hereafter Paper~I and II, respectively), showing that quite
different excitation conditions can occur in HH objects: while in HH43 the
bulk of the cooling occurs in H$_2$ lines, the spectra of the other three
objects are dominated by iron lines. In HH43, the H$_2$ emission has been
successfully fitted by means of high velocity, C-type shock models (Flower
et al. 2003), whereas, in the other three cases, the strong ionic emission
testifies in favour of a dissociative component in the shock structure.
Here we present the observations obtained in three other regions, namely
HH24-26, HH72 and HH320/321 (BHR71).  All the spectra exhibit copious
H$_2$ rovibrational emission, together, in some cases, with a fainter
atomic and ionic component. Therefore, they represent valid tests
of the ability of current shock models to predict simultaneously molecular
and atomic/ionic emission and an opportunity to check the validity
of the underlying assumptions of the models.

The structure of the paper is the following: we present the targetted
regions in Sect. 2 and then describe the observations and the results
obtained (Sect. 3). In Sect. 4, we derive the physical parameters of the
emitting gas and model the observed emission, assuming shock excitation.
Sect. 5 summarizes our conclusions.

\section{The investigated regions}
\subsection{The HH24-26 region}


Located in the L1630 Orion dark cloud (d $\approx$ 400 pc, Anthony-Twarog
1982), the HH24-26 region is an active site of star formation, rich in
young stellar objects in different evolutionary stages. Two low
luminosity, Class 0 protostars (Andr\'e et al. 2000), HH24MMS (Chini et
al. 1993, $\alpha_{2000}$=05$^{\rm h}$46$^{\rm m}$06.7$^{\rm s}$,
$\delta_{2000}$=$-$00$\degr$10$\arcmin$40.8$\arcsec$, 
{\it l$^{II}$}=205.49,{\it b$^{II}$}=$-$14.57, L$_{bol}$=5L$_{\odot}$) 
and HH25MMS (Bontemps et al.1995, $\alpha_{2000}$=05$^{\rm h}$46$^{\rm
m}$06.7$^{\rm s}$, $\delta_{2000}$=$-$00$\degr$13$\arcmin$24.7$\arcsec$,
{\it l$^{II}$}=205.53,{\it b$^{II}$}=$-$14.60,
L$_{bol}$=6L$_{\odot}$) drive distinct compact jets, traced by means of
both the 2.12 $\mu$m H$_2$ line and CO mm transitions (Gibb $\&$ Davis
1998; Bontemps et al. 1996), while the Class I protostar HH26IR (Davis et
al. 1997, $\alpha_{2000}$=05$^{\rm h}$46$^{\rm m}$03.9$^{\rm s}$,
$\delta_{2000}$=$-$00$\degr$14$\arcmin$52.5$\arcsec$,
{\it l$^{II}$}=205.54,
{\it b$^{II}$}=$-$14.62, L$_{bol}$=28.8L$_{\odot}$) gives rise to a
more extended molecular outflow
(Gibb $\&$ Heaton 1993).  Imaging in the 2.12 $\mu$m H$_2$ rovibrational
line (Davis et al. 1997, see Figure 1) reveals a string of shocked H$_2$
knots along the axis of the HH25MMS and HH26IR outflows. Herbig-Haro
objects\footnote{In the following we will refer to HH objects as the
optically visible condensations.} (HH24, HH25 and HH26) are associated
with each of these three sources. Proper motions were measured in HH25 and
26 by Chrysostomou et al. (2000), who found velocities ranging from less
than 70 up to 200 km s$^{-1}$; the lowest value pertains to HH26A, which
should be a stationary shock region, and the largest one to HH26C in the
working surface of a bow shock. Far infrared spectra of the region were
obtained by Benedettini et al. (2000), who interpreted the observed
emission (from {[{\ion{O}{i}}]}, CO, H$_2$O and H$_2$ lines) as due
to a mixture of both C- and J-type shocks. Recently, Fabry-Perot
observations have demonstrated the existence of elongated H$_2$ emission,
on scales of a few arcsec, from HH26IR and HH25MMS, which appears
associated with the base of the larger scale jets (Davis et al. 2002).

\subsection{The HH72 region (L1660)}

The L1660 region is a bright-rimmed core illuminated by the young OB
cluster NGC2362 in Vela (Reipurth $\&$ Graham 1988), at a distance of
$\approx$\,1500 pc (Hilton $\&$ Lahulla 1995). An E-W, poorly collimated
outflow (Schwartz, Gee $\&$ Haung, 1988) is present in the region, possibly driven
by the intermediate luminosity IRAS source 07180-2356 (Reipurth et al. 1993,
$\alpha_{2000}$= 07$^{\rm h}$20$^{\rm m}$10.3$^{\rm s}$,
$\delta_{2000}$=$-$24$\degr$02$\arcmin$24$\arcsec$, 
{\it l$^{II}$}=237.52,{\it b$^{II}$}=$-$4.83, L$_{bol}$ $\approx$
170L$_{\odot}$). Figure 2 shows the SOFI 2.12 $\mu$m image of the region,
where, together with the chain of HH objects (HH72 A-C, Reipurth $\&$
Graham 1988), several pure H$_2$ features are recognizable (HH72 D-Z). A
jet of a few arcseconds has been recently observed close by the infrared
source HH72-IRS (Davis et al. 2002, see Fig.2).

\subsection{The BHR71 region}
Figure 3 shows the H$_2$ 2.12$\mu$m image by Bourke (2001) of the Bok
globule BHR71 (Bourke, Hyland $\&$ Robinson 1995), located at $\approx$
200 pc (Seidensticker $\&$ Schmidt-Kaler 1989). An embedded binary
protostellar system, separated by $\approx$ 3400 AU, has been found in the
region (Bourke 2001). One of the two sources has been identified as a
Class 0 protostar (IRS 1: $\alpha_{2000}$=12$^{\rm h}$01$^{\rm
m}$36.6$^{\rm s}$, $\delta_{2000}$=$-$65$\degr$08$\arcmin$48.2$\arcsec$,
{\it l$^{II}$}=297.73, {\it b$^{II}$}=$-$2.78,
L$_{bol}$$\approx$9L$_{\odot}$) by means of 1.3 mm observations (Bourke et
al. 1997); ISO mid-IR maps led to the discovery of the possibly more
evolved source IRS 2 (Bourke 2001, $\alpha_{2000}$=12$^{\rm h}$01$^{\rm
m}$34.0$^{\rm s}$, $\delta_{2000}$=$-$65$\degr$08$\arcmin$44.3$\arcsec$,
{\it l$^{II}$}=297.72, {\it b$^{II}$}=$-$2.78). A
recent study by Lemaire et al. (2002) suggests the presence of a
circumstellar disk around this object. A large scale CO outflow, where a
shock enhanced chemistry has been revealed (Garay et al. 1998), is driven
by IRS1 along the North-South direction, while IRS2 gives rise to a more
compact outflow with an inclination $\approx$$-$36$\degr$. Optical
[{\ion{S}{ii}}] images revealed HH associations (HH320 and HH321)
along both the CO outflows (Corporon $\&$ Reipurth, 1997); these should be
at a very low degree of excitation, as demonstrated by the lack of
emission in the H$\alpha$ line. Faint H$_2$ emission features were found
by Bourke (2001).


\section{Observations and Results}
\addtocounter{figure}{+3}
\begin{figure*}[!ht]
\centering
\includegraphics[width=15cm]{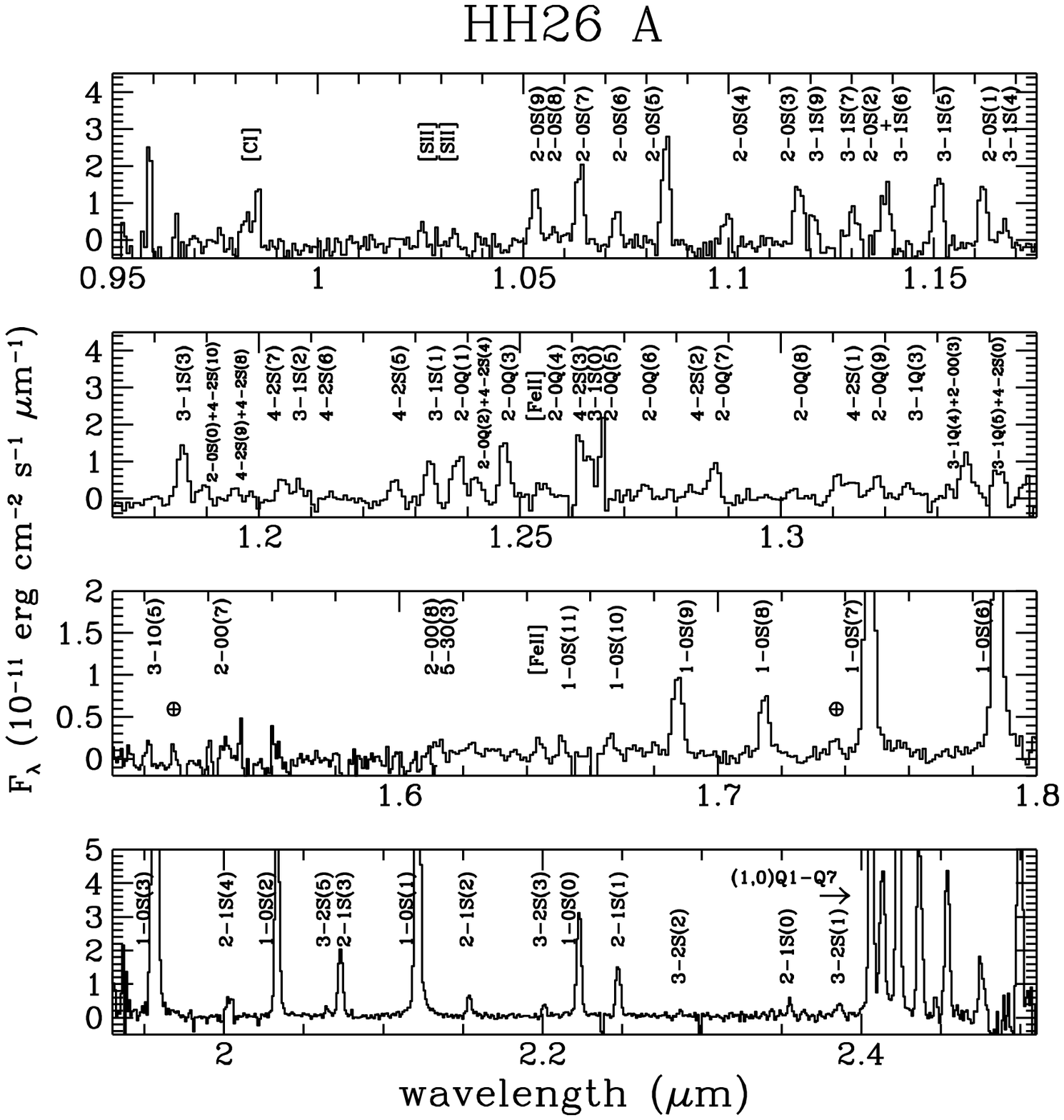}
\caption{Spectrum between 0.95 and 2.5 $\mu$m of HH26A.}
\end{figure*}


\begin{figure*}[!ht]
\centering
\includegraphics[width=15cm]{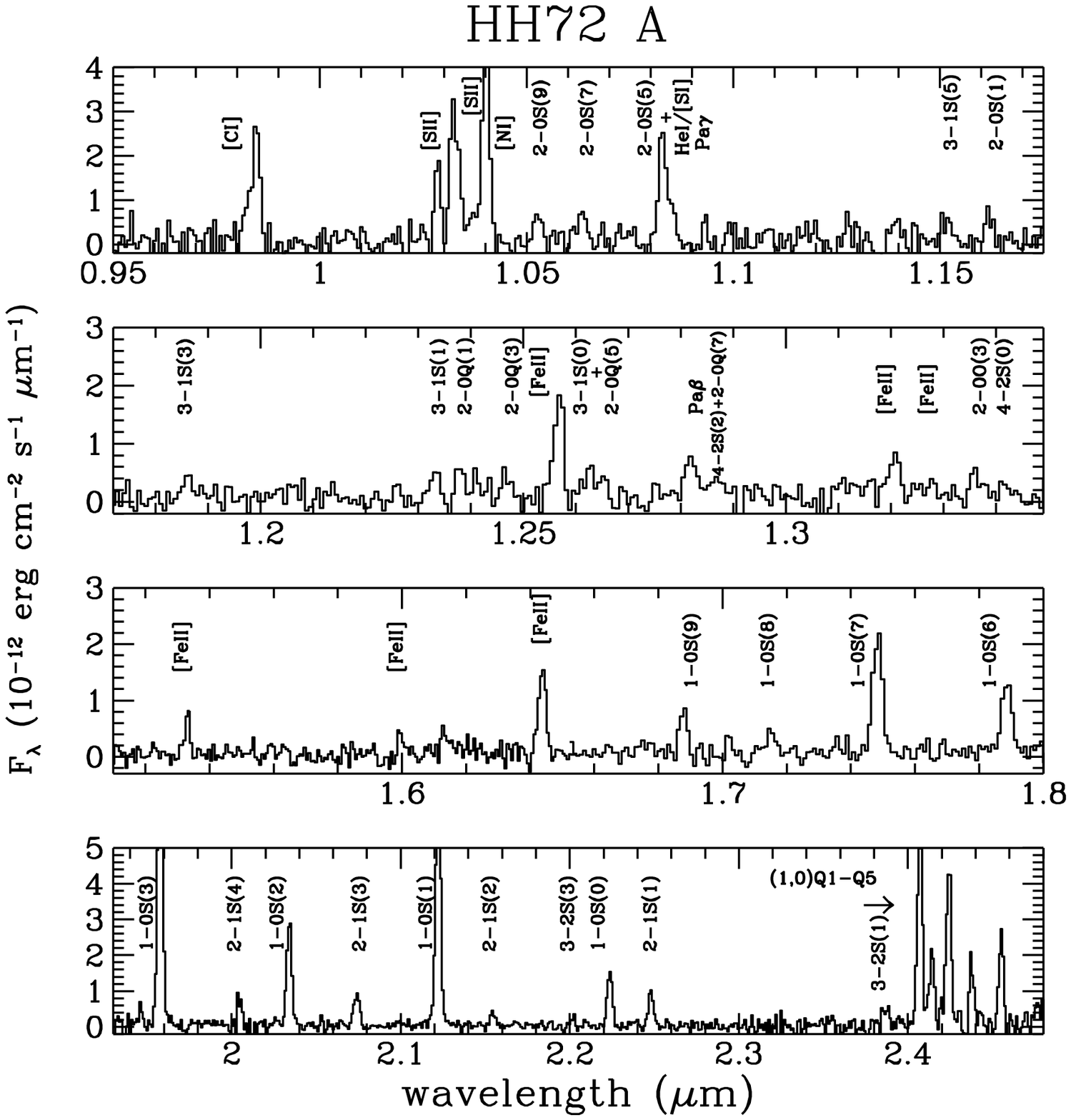}
\caption{Spectrum between 0.95 and 2.5 $\mu$m of HH72A. }
\end{figure*}

\begin{figure*}
\centering
\includegraphics[width=15cm]{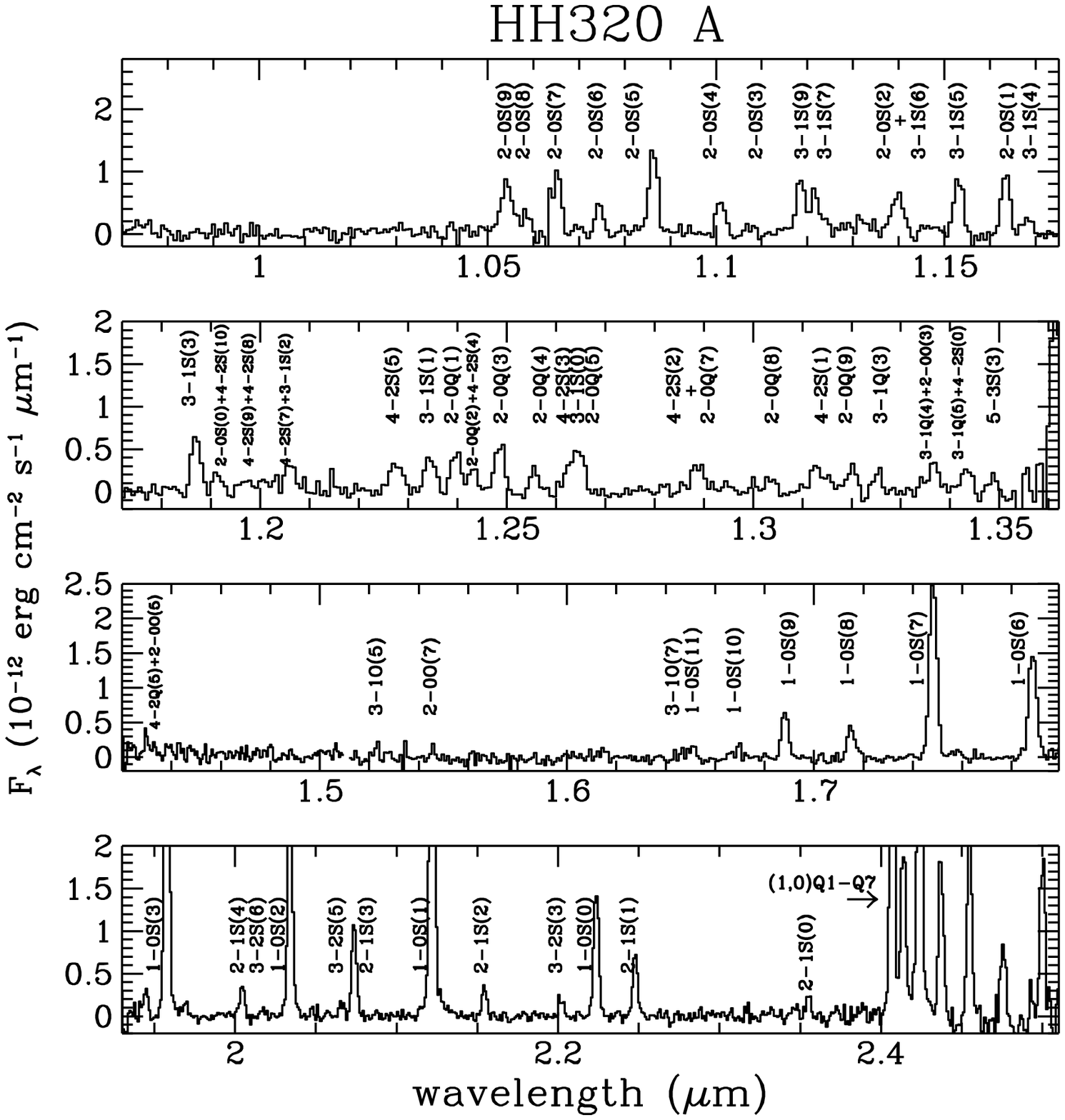}
\caption{Spectrum between 0.95 and 2.5 $\mu$m of HH320A.}
\end{figure*}

The observations were carried out during two runs (January 2001 and 2002)
with SOFI (Lidman \& Cuby 1999) at ESO-NTT 3.5m telescope (La Silla). Long
slit spectroscopy was obtained through the blue (0.95-1.64 $\mu$m) and the
red (1.53-2.52 $\mu$m)  grisms in the low resolution mode
($\lambda$/$\Delta\lambda$ $\approx$ 600 with the 1 $\times$ 290 arcsec
slit). To cover most of the H$_2$ emission knots present in the HH24-26
region we used five different slit orientations, which are depicted in
Figure 1. In the HH72 region, we adopted four slits to observe the knots
both in the blueshifted (knots L-B) and redshifted (knots Y
and Z) lobes of the outflow (Figure 2), while the two jets present in
BHR71 were investigated with three slits (see Figure 3), covering both the
Herbig-Haro objects 320/321 and the fainter H$_2$ emission features
recognizable in the 2.12 $\mu$m image by Bourke (2001). The observations
were performed by nodding and jittering the telescope, keeping the target
along the slit in the usual ABB'A' mode, with a total integration time of
1200 s. The data reduction and calibration were performed by using the
IRAF package. Each observation was flat fielded and sky subtracted, while
to remove the atmospheric spectral response a couple of telluric O-type
stars were observed before and after each on-source observation. Each
target spectrum was then divided by the telluric spectrum, corrected by
the blackbody function at the temperature of the star, and by a few
intrinsic hydrogen absorption lines. Wavelength calibration was obtained
from the spectrum of a Xenon-Argon lamp; the estimated wavelength error is
smaller than the spectral resolution element (2 pixels $\approx$ 20 \AA),
which corresponds to a velocity of about 300 km s$^{-1}$. No evidence of
shifts larger than this limit has been found. Flux calibration (associated
uncertainty 10$\%$) was obtained from the observation of some photometric standard 
stars from the Carter $\&$ Meadow (1995) catalogue. Ratios between lines 
lying in the blue and red parts of the spectrum, are affected by a 10$\%$ 
uncertainty, due to the different instrumental response in the two segments. Such error has
been estimated  by using a line (e.g. [{\fe}] 1.64$\mu$m)
present in both parts. As an example of our observations
we show in Figs. 4, 5 and 6 the spectra of HH26A, HH72A and HH320A, which
are among the knots richest in lines; they are all dominated by H$_2$
emission in form of rovibrational lines coming from levels with
vibrational quantum number up to 5. A relatively faint atomic
and ionic emission is also observed: in addition to the bright [\ci]~ doublet
at 0.98 $\mu$m, a
few lines of [\fe], [\s], and [\n] were detected in the most prominent
knots. Tables 1a-c, 2a-b and 3a-b (which are available only in the electronic
version of the paper), list all the detected lines, along with
their spectral identification and the vacuum wavelength. Some knots (e.g.
HH26A,C) were intersected by two slits, but we did not find a significant
variation in the line ratios along the different directions, although the
emission is generally brighter along the jet axis. To the latter
correspond the fluxes reported in the Tables. The integrated fluxes were
obtained by fitting the line shape with a single (or double in case of
blending) gaussian profile. The reported uncertainty derives from
the rms of the baseline, multiplied by the width of the instrumental
profile (30 \AA).\\
The H$_{2}$ 1-0 S(1) 2.12$\mu$m line fluxes observed in HH24-26 and in
HH72 are systematically lower, by a factor between 1.5 and 10, than the
values derived from imaging by Davis et al. (1997).  The discrepancies are
attributable to the spatial extent of the objects observed, relative to
our 1$^{\prime\prime}$ slit width; this explanation is supported by the
fact that the largest differences occur in those knots, such as HH26D,
that we observed along their shortest axis (see Figure 1).\\
From inspection of Tables 1a-1c, we see that, along the HH26 flow, copious
H$_2$ emission is detected only close to the driving source (HH26A and C),
while in all the other knots faint and weakly excited H$_2$ lines are
revealed. The opposite situation exists along the HH72 jet (Tables
2a-2b), where the richest H$_2$ spectra are emitted by the knots at the
apex of the blueshifted lobe (HH72A and B).  Only H$_2$ emission has been
detected toward the BHR71 jets (Tables 3a-3b). Particularly remarkable is
the lack of the [{\ci}] doublet, which, in addition to the present
observations, has been detected in all the HH objects previously
investigated (Paper~I,~II).  We also note that, compared with the other
Herbig-Haro objects in BHR71, HH321B has a significantly weaker spectrum,
probably because the emission peak was badly covered by the slit.\\ To
complement the SOFI observations, we obtained in April 2002 the 3-5 $\mu$m
spectra of the HH72 and BHR71 regions with ISAAC at the VLT (Cuby et al.
2003). In this spectral range, several pure rotational lines of H$_2$ with
high excitation energy (up to $\approx$ 40\,000 K)  are located, which can
be used to determine the density of the molecular gas (Giannini et al.
2001). Unfortunately, the poor atmospheric conditions prevented us to
obtain high quality data at wavelengths larger than 4 $\mu$m, where both
the thermal emission and the presence of strong water absorption bands
critically affect the reliability of the observations. We used the
1$\times$120 arcsec slit ($\lambda/\Delta\lambda \approx$ 2000), pointing
at the Herbig-Haro objects 72 A/B (P.A. 84$^{\circ}$, i.e. slit I in Fig.2) 
and 320A/B (P.A.320$^{\circ}$, i.e. slit I in Fig.3). The observational strategy 
and the procedure adopted for the data reduction are similar to those reported by 
Giannini et al.(2001). We detected two H$_2$ lines: the 1-0 O(5) at 3.235 $\mu$m in HH72B
and the 1-0 S(13) at 3.847 $\mu$m in HH72B and HH320A (Tables 2a, 3a). The
extracted spectra are shown in Figure 7. The 1-0O(5) line appears narrower
of about factor of 2 than the instrumental profile; however, since its trace 
in the bidimensional image of the spectrum is clearly recognizable, we
consider the line to be a genuine detection. The estimated fluxes were used in
conjunction with the SOFI data.  No intercalibration correction was
applied to the ISAAC and SOFI lines, but the consistency of these two data
sets is supported by the fact that the column densities of lines coming
from the same upper level are equal to within the error bars (see Sect.
4.2).

\section{Analysis and Discussion}

\subsection{$A_\mathrm{V}$ determination}

The line lists, reported in Tables 1-3, clearly suggest that some
differences in the H$_2$ emission exist throughout all the observed knots.
In particular, lines coming from the highest excited levels ($\nu \ge$ 3)
are observed mainly in condensations visible also in the optical (e.g.
HH26A/C, HH72A/B, HH320/321), while low excitation conditions are
recognizable, with few exceptions, in the other knots (e.g. HH25,
HH72E-Z). Since the most excited lines are located mainly in the blue
part of the spectrum (i.e. at wavelengths shorter than 1.4 $\mu$m), their
absence in the pure infrared knots could be interpreted as being due to a
higher visual extinction, without necessarily implying lower excitation
conditions. To investigate this crucial point, we have firstly estimated
the visual extinction in the various knots by using pairs of transitions
coming from the same upper level, for which the difference between the
observed and the theoretical flux ratio is a function only of $A_\mathrm{V}$ (cf.
Gredel 1994). The accuracy of this method relies strongly on the presence
of pairs of bright lines at significantly different wavelengths which are
not affected severely by observational problems. In practice, sufficient
numbers of such pairs of lines are present in the spectra of only the
brightest, optical knots. For this reason, the values of $A_\mathrm{V}$ derived in
the brightest objects were applied also to the nearby knots, where
insufficient lines are present to allow a specific determination. Given the
large number of molecular hydrogen lines, the criteria cited above are
satisfied by several pairs of lines coming from the same upper level (e.g.
of the kind 2-0 S(i)/2-0 Q(i+2) or 3-1 S(i)/3-2 S(i)). By using all the
available ratios, we were able to estimate the visual extinction with an
uncertainty of 1-2 mag. Adopting the reddening law of Rieke \& Lebofsky
(1985), we found $A_\mathrm{V}$ up to 8 mag in the HH24-26 and HH72 flows,
while the visual extinction in the BHR71 region is small ($A_\mathrm{V}$$\le$2\,mag),
however well compatible with the $A_\mathrm{V}$ values ($<$ 3 mag)
estimated towards the galactic plane direction (e.g. Hakkila et al. 1997).
The values of $A_\mathrm{V}$ for individual sources are given in Figs. 8-13.
These results indicate that the differential extinction only marginally
biases the observations in all the knots except those having Av=8\,mag (HH25B,
HH72J-Z).
Thus the presence or absence of highly excited H$_2$ lines in the spectra
reflects in almost all cases the excitation conditions of the molecular
gas. In this respect, the fact that we detect no lines 
with $\nu$$\ge$6 indicates that fluorescence mechanisms do not play an 
important role in the excitation (Black $\&$ van Dishoeck 1987).\\
An independent determination of $A_\mathrm{V}$ could be obtained from the [{\fe}]
lines at 1.257 and 1.644 $\mu$m. However, in our objects, these lines are
generally detected with a poor signal-to-noise ratio, with the only
exception being HH72A. For this object, we derive
$A_\mathrm{V}$$_{[{\fe}]}$=4.7$\pm$1.5 mag, a value higher than the negligible
extinction indicated for this object by the H$_2$ lines. The possible
implications of this result will be discussed below.

\begin{figure}[h]
  \includegraphics[width=8cm]{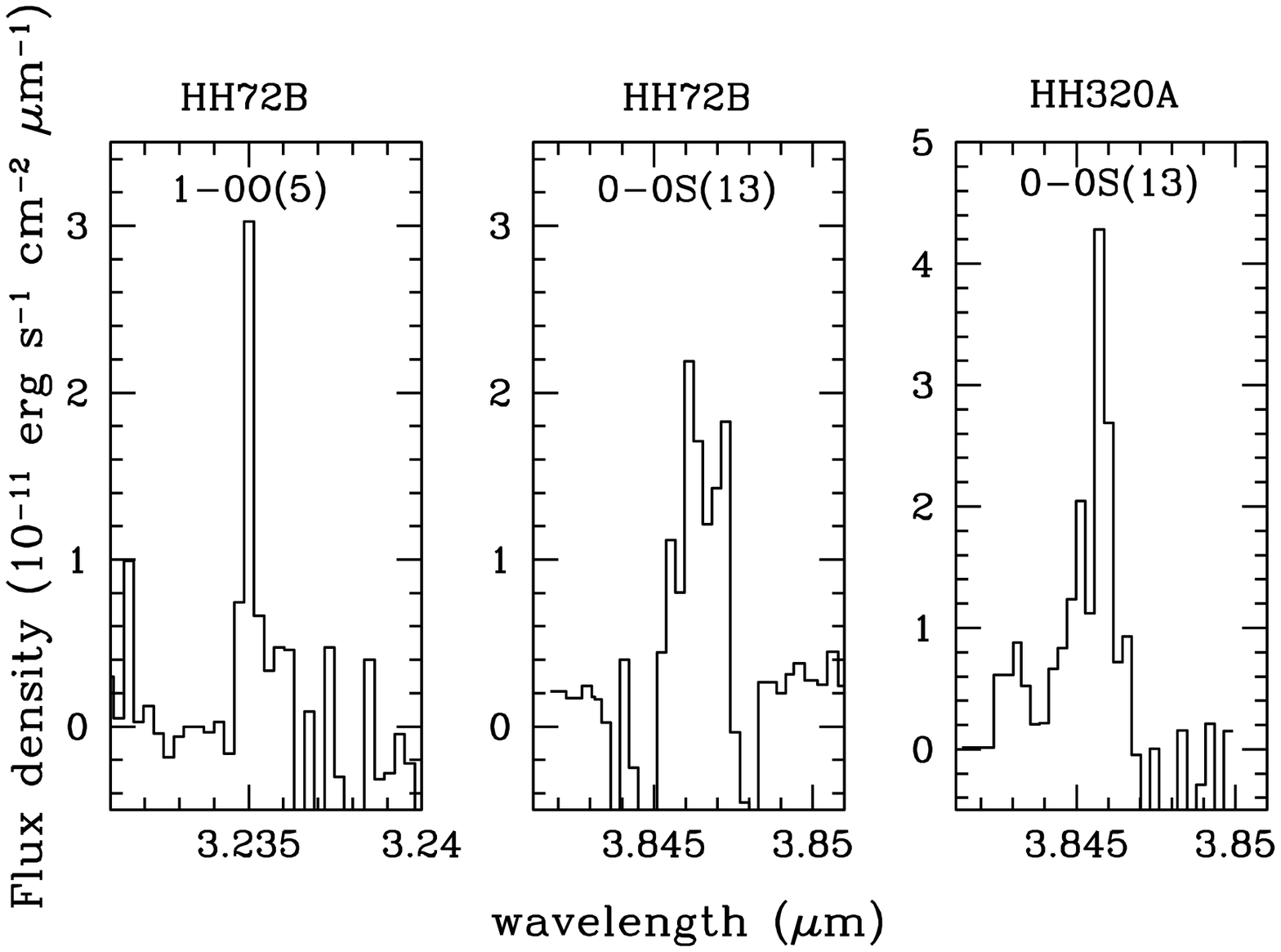}
  \caption{H$_2$ lines observed with ISAAC in HH72B and HH320A.}
\end{figure}
\subsection{\h~ emission}

\begin{figure}
\includegraphics[width=8.5cm]{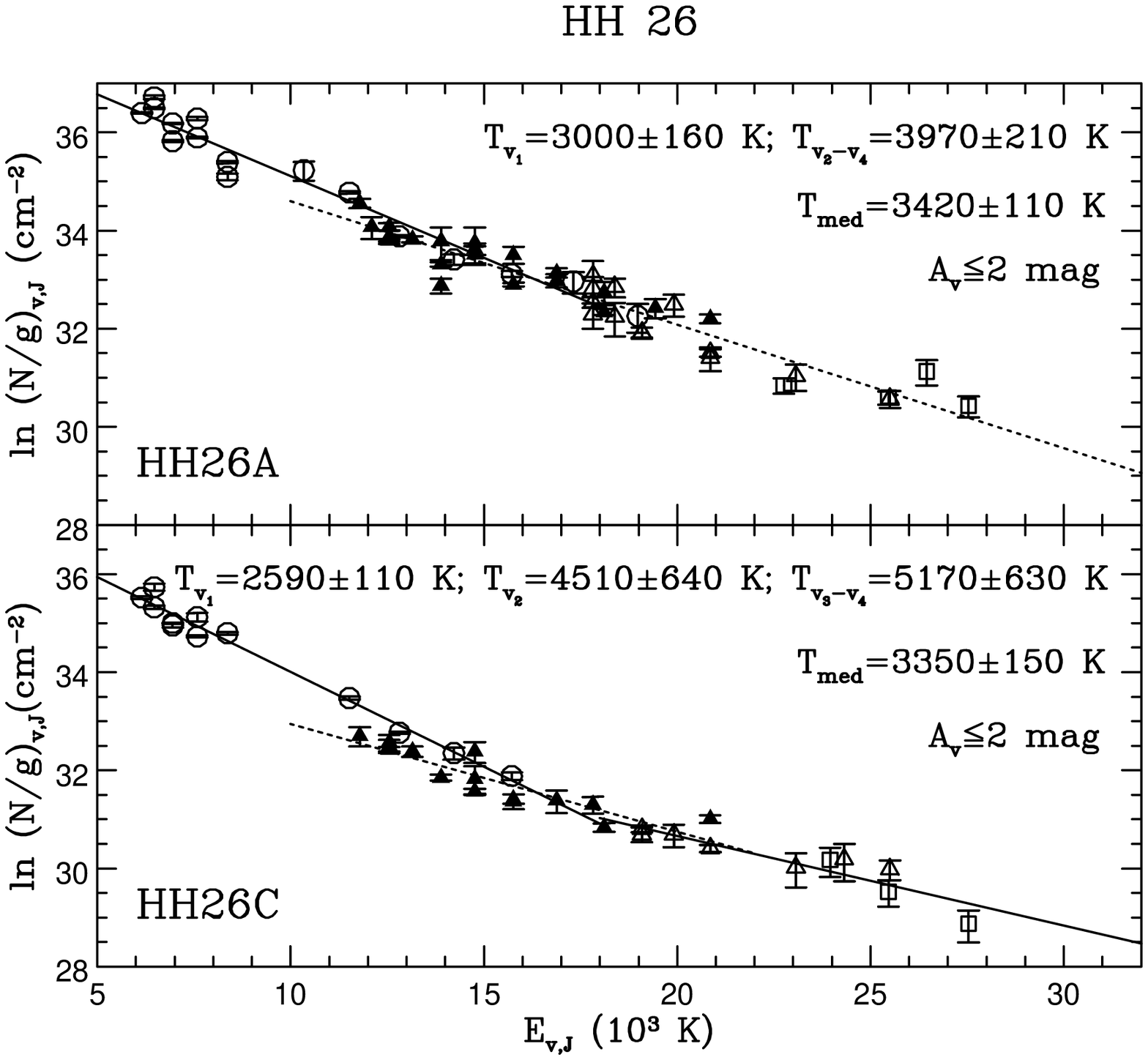}
\caption{Excitation diagram for HH26 A and B. Lines coming from different
vibrational levels are indicated with
different symbols: filled circles, open circles, filled triangles, open
triangles, filled squares and open
squares indicate lines coming from levels from 0 to 5, respectively. The
straight line(s) represents the best fit(s)
through the data, at the temperature(s) reported in top right angle. The
derived value of the extinction is also
indicated.}
\end{figure}

\begin{figure}
\includegraphics[width=8.5cm]{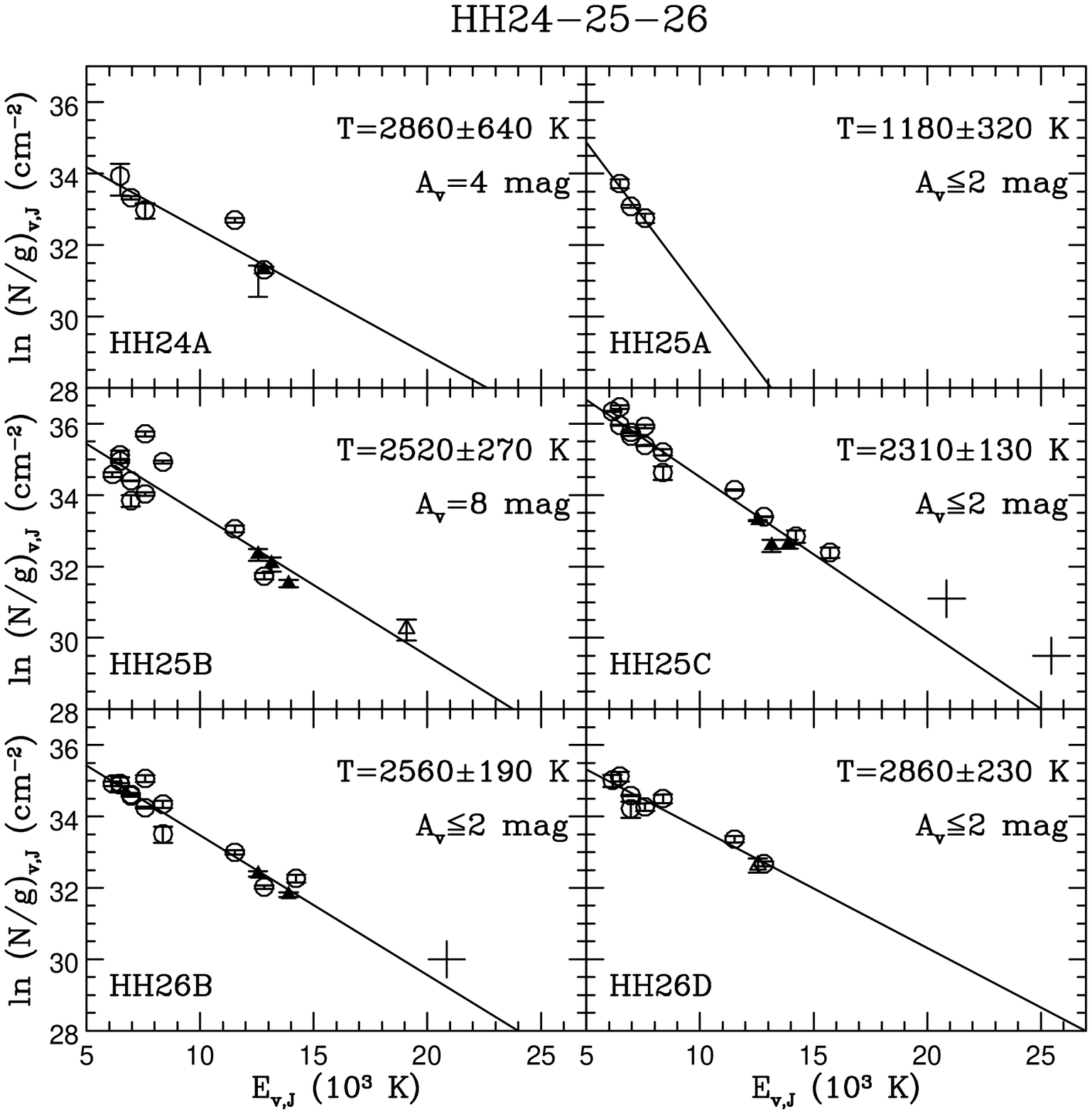}
\caption{As in Figure 8 for the pure H$_2$ knots in the HH24-26 region. The
two crosses in the HH25C panel indicate the positions
which would correspond to the 3-1 S(5) and 4-2 S(5) lines by assuming the
same ratios with the 1-0 S(1) line as in the HH26C
case (see text).}
\end{figure}

\begin{figure}
\includegraphics[width=8.5cm]{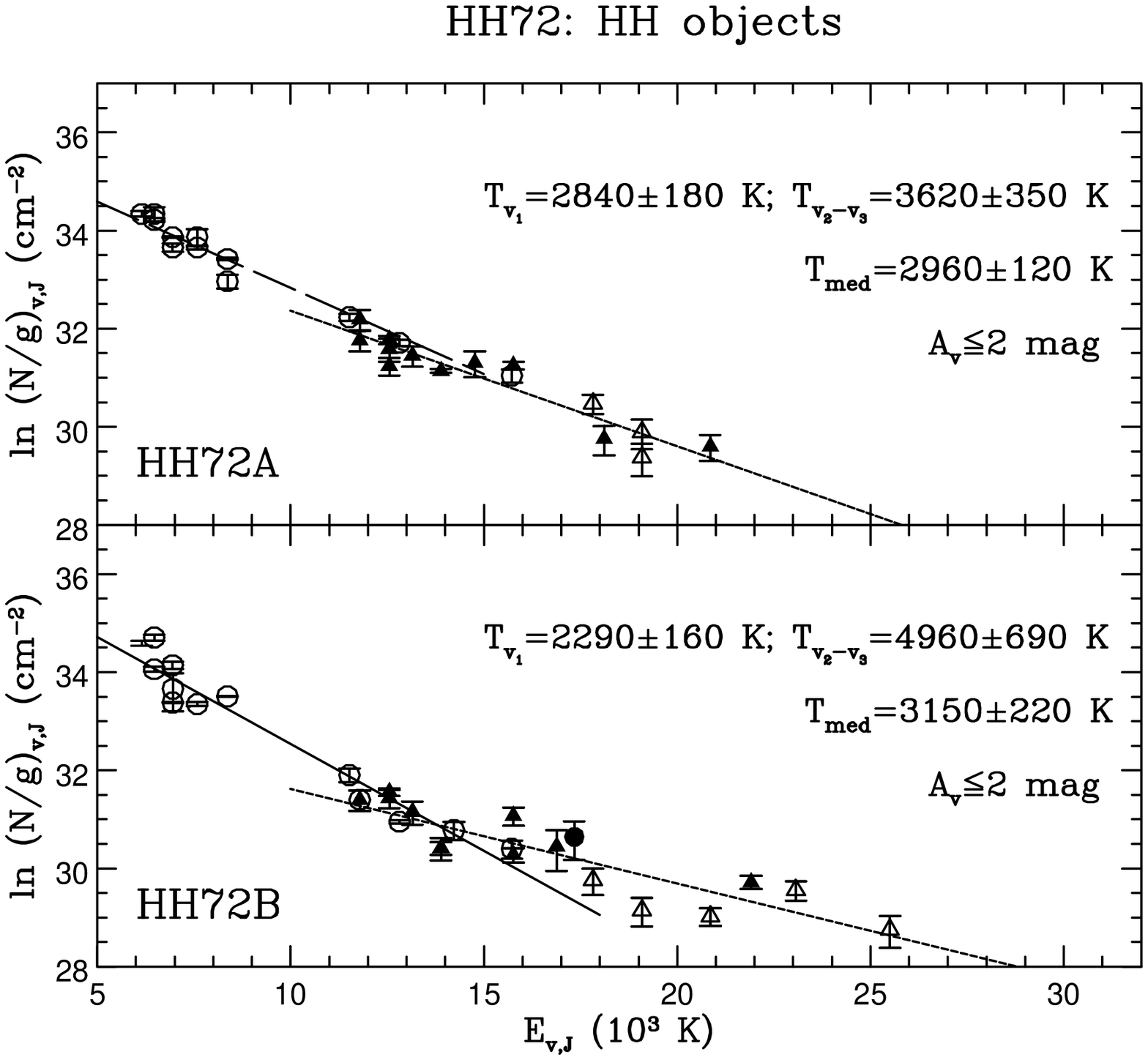}
\caption{As in Figure 8 for HH72 A and B.}
\end{figure}

\begin{figure}
\includegraphics[width=8.5cm]{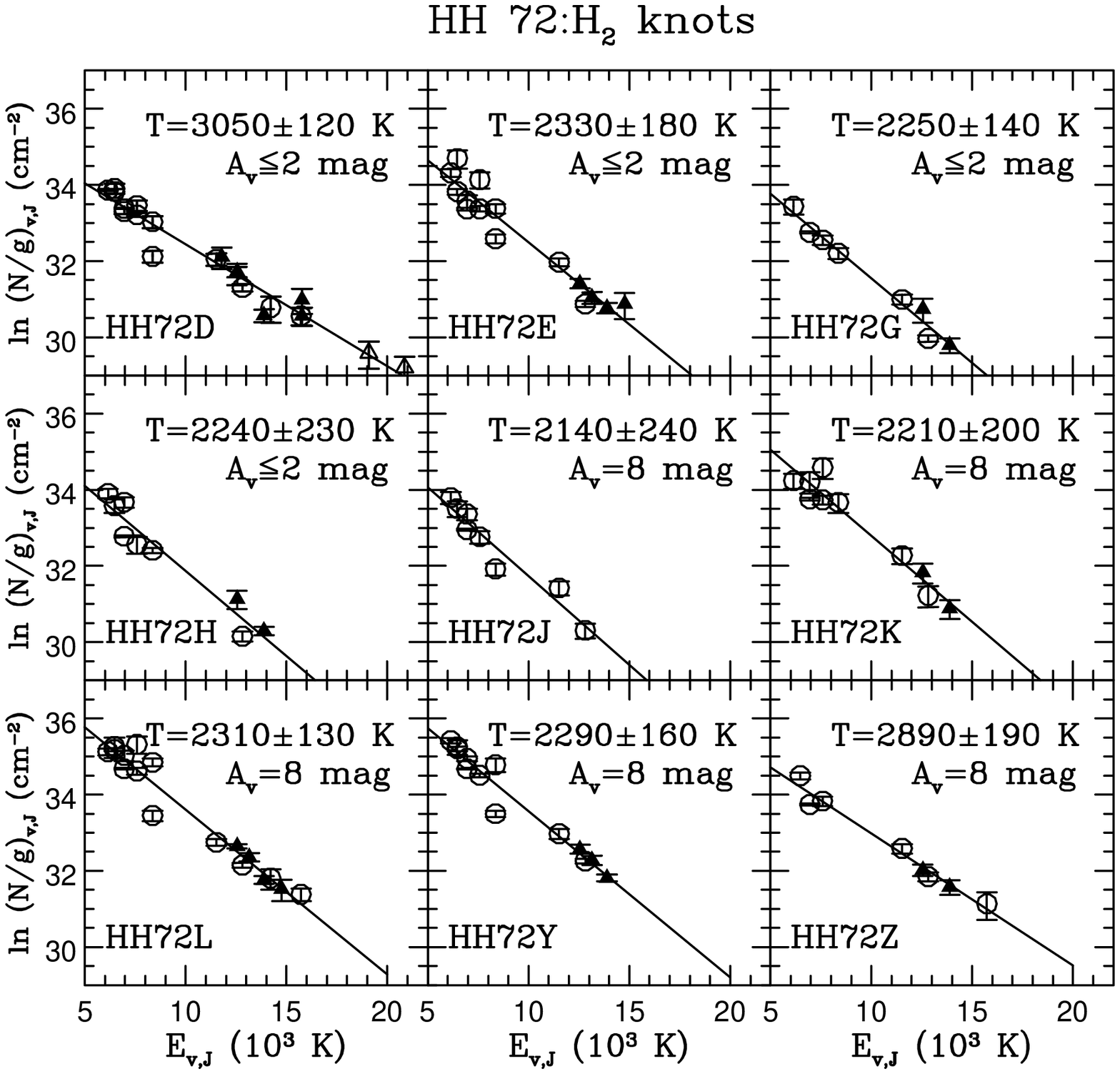}
\caption{As in Figure 8 for the pure H$_2$ knots in the HH72 region.}
\end{figure}

\begin{figure}
\includegraphics[width=8.5cm]{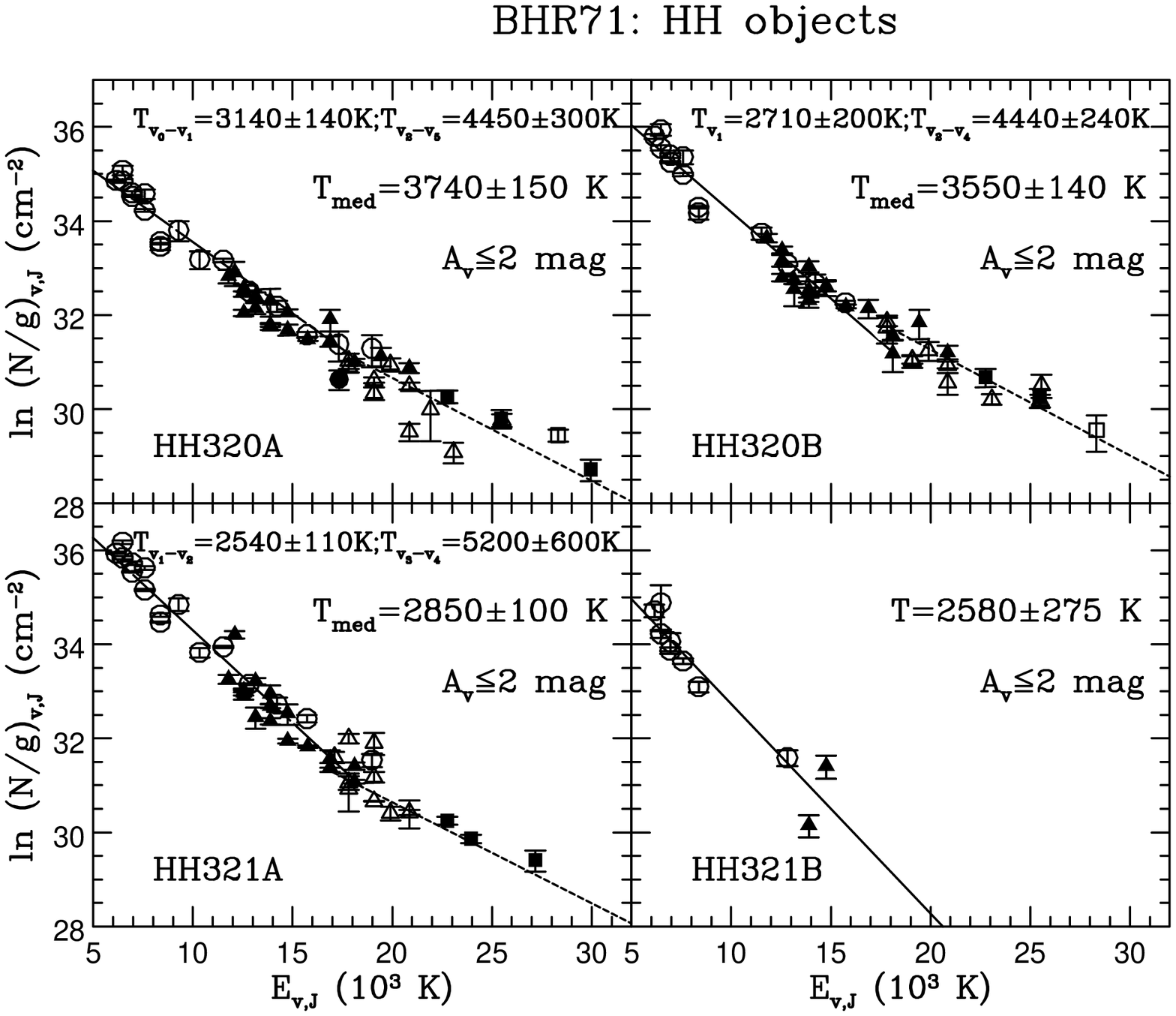}
\caption{As in Figure 8 for HH 320A/B and HH321A/B.}
\end{figure}

\begin{figure}
\includegraphics[width=8.5cm]{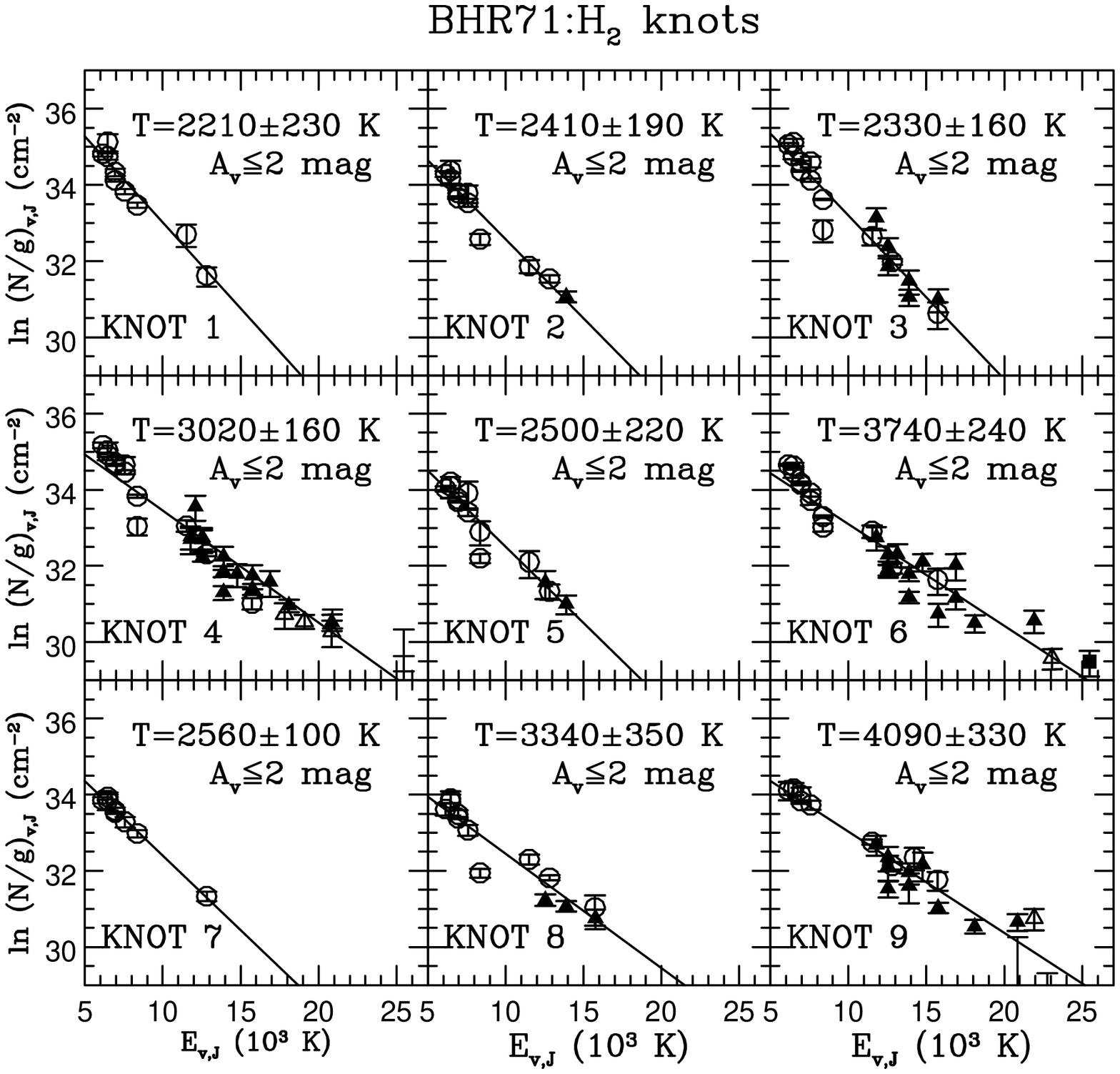}
\caption{As in Figure 8 for the pure H$_2$ knots in the BHR71 region.}
\end{figure}

In order to derive an estimate of the temperature along the flows, we have
used the excitation diagrams (Boltzmann plots) for the H$_2$ lines
(Figures 8-13), plotting the extinction-corrected column densities of the
rovibrational levels, divided by the statistical weights, against their
excitation energies. 
In each plot, lines coming from different vibrational
states (from $\nu$=0 to $\nu$=5) are depicted by different symbols. To
minimize the uncertainties, we considered only unblended lines with a
signal-to-noise ratio greater than 3. Any significant deviation of the ortho:para
ratio from its statistical value of 3 would be apparent as a 
misalignment of the ortho and para data points in the rotation diagrams.
No such deviations are observed: see below and Figs. 8-13.\\
If the gas is excited at a
single temperature, a straight line can be fitted through the data
points: this seems to be the case only where lines with excitation
energies lower than typically 15\,000 K are detected, i.e. in all the
molecular hydrogen condensations and few Herbig-Haro objects (such as
HH321B and HH25A).  Here a single temperature typically between
2000 and 3000 K can account for the observed excitation diagrams.\\
Other Herbig-Haro objects (HH26A, HH72A/B, HH320A/B and HH321A) and HH26C,
which show a richer H$_2$ spectrum, deserve a more realistic approach,
which consists of relaxing the assumption of a single excitation
temperature and fitting the various vibrational series separately.
Although the average temperature remains quite similar to that measured in
the molecular condensations, deviations from the thermal equilibrium are
evident, with a spread from 2000 K to more than 5000 K. Similar conditions
have been found in two out of the four HH objects we investigated in
previous studies, namely HH43 (Paper~I) and HH120 (Paper~II).\\
Since the lack of a temperature stratification is deduced only when
the highly excited lines
(with $\nu\ge$ 3) remain undetected, it is crucial to understand whether,
in these cases, the observations reflects an intrinsic property
of the gas or is attributable to the instrumental detection limit.

In order to investigate this point, we have estimated the fluxes expected
for the $\nu \ge$ 3 lines, both from a single temperature gas  and in the
case of
a temperature stratification. As a test case, we considered the spectrum
of HH25C, which is a bright knot with a negligible extinction,
observed in only $\nu$=1,2 lines, which can be fitted by a single
temperature $\approx$~2300 K (Figure 9, middle right panel). By
extrapolating the straight line up to energies $\approx$ 25\,000 K, we can
evaluate the fluxes of the $\nu$=3,4 lines that would be expected in
thermal equilibrium: we find that all the $\nu$=3,4 lines are definitely
below the detection limit. We have estimated the fluxes of the same
lines, in case of a temperature stratification, by assuming that their
ratios with the 1-0 S(1) line are the same as those measured in HH26C (a
knot with brightness and extinction comparable to HH25C, but
showing a spectrum richer of $\nu$=3,4 lines). In Figure 9, the
two crosses indicate the corresponding column densities of the 3-1 S(5)
and 4-2 S(5) lines: both these points lie above the 3$\sigma$
detection limit. Thus we conclude that the absence of these lines in the
spectrum of HH25C points to excitation conditions different from HH26C.\\
This analysis has been repeated for other bright knots (such as HH72D
and knots 4 and 9 in BHR71), yielding the same conclusion as in the case
of HH25C. Thus temperatures rarely exceeding
3000K appear to be a proper feature of the pure infrared condensations,
while in HH objects significant temperature gradients occur. This
tentative conclusion needs to be confirmed by enlarging the observational
database.

\subsection{Atomic and Ionic emission}

\begin{figure}
\includegraphics[width=8.5cm]{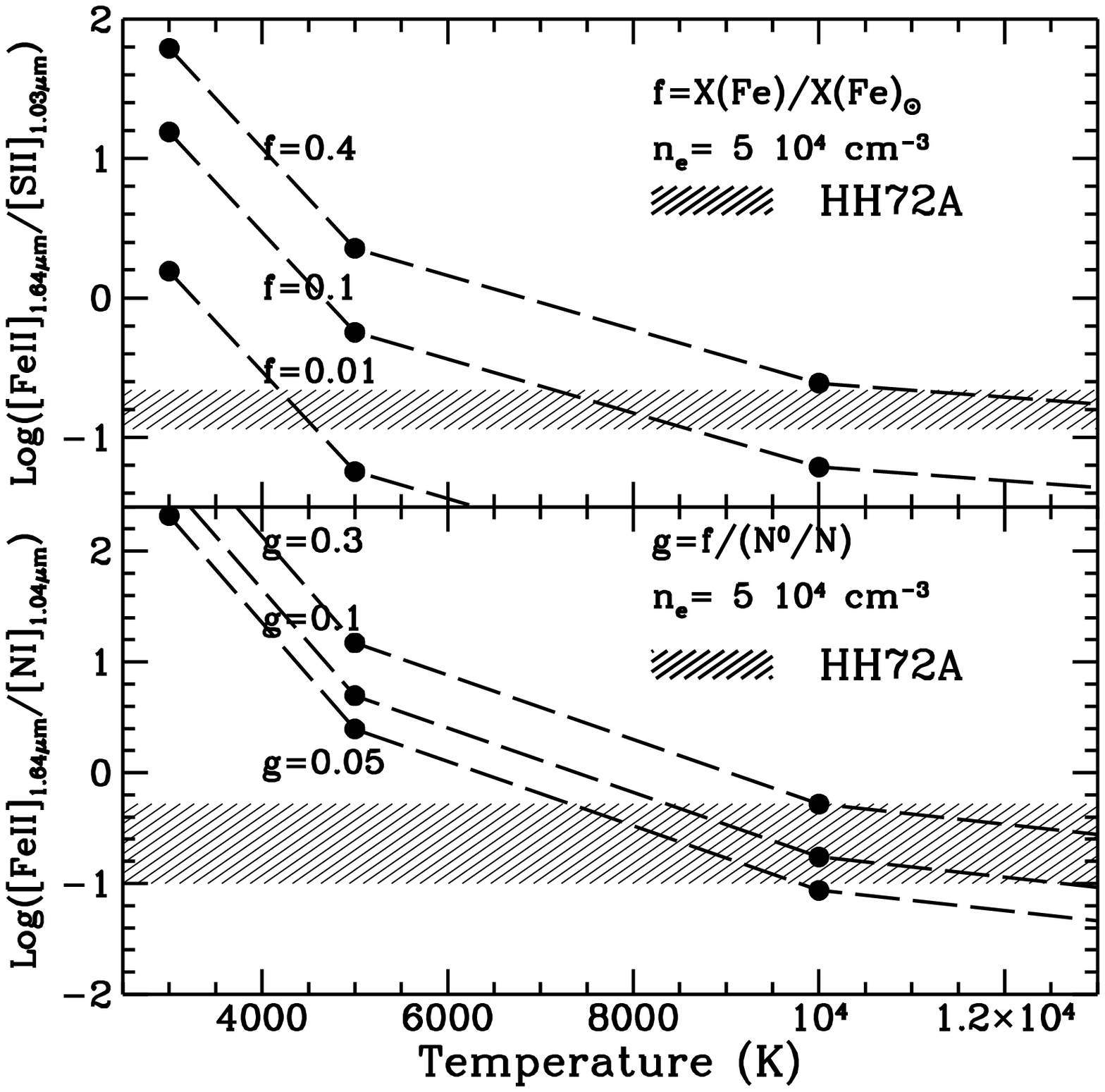}
\caption{Top panel: [{\fe}]~1.644~$\mu$m/[{\s}]~1.03~$\mu$m as function of
the temperature, for an electron
density of 5 $\times$ 10$^4$ cm$^{-3}$. The curves refer to different
assumptions on the iron abundance.
The observed ratio in HH72A is drawn as a hatched area. Bottom
panel: as in the top panel
for the ratio [{\fe}]~1.644~$\mu$m/[{\n}]~1.04~$\mu$m. The parameter g is
the ratio between the iron abundance
and the fraction of neutral nitrogen.}
\end{figure}

As we have already pointed out, faint atomic and ionic lines are
detected in only
four of the objects investigated (HH24A, HH26A and HH72A/B). We first
attempt to analyze such emission without making assumptions regarding the
underlying excitation mechanism, using simple NLTE models of the
equilibrium level populations. We then present the results obtained by
modelling the same emission, assuming shock excitation (Sect. 4.4.6).

Following the procedure described in Paper~II, we determine the electron
density from the [{\fe}] 1.644~$\mu$m/1.600~$\mu$m ratio, obtaining 
$n_\mathrm{e}$ of
approximately 5 $\times$ 10$^4$ cm$^{-3}$ for HH24A and HH72A, a value
which can be considered also as a 3$\sigma$ upper limit for HH26A. Since
all the transitions of [{\fe}] detected come from levels with quite
similar excitation energies, their relative intensities are insensitive to
the temperature. However, some constraints on the conditions in the
ionized gas can be derived by using lines from different species, such as
the [{\s}] and the [{\n}] multiplets at 1.03~$\mu$m and 1.04~$\mu$m. In
practice, this can be done for HH72A, the only object where these lines
are all detected with a sufficiently high signal-to-noise ratio.\\ From
the [{\fe}]~1.644~$\mu$m/[{\s}]~1.03~$\mu$m ratio we derive the fraction of
iron
in the gas phase, a parameter regulated in HH objects by the shock
efficiency in destroying the cores of the dust grains in which most of the
iron is to be found (cf. May et al. 2000).  To this end, we plot, in the
upper panel of Figure 14, the expected
[{\fe}]~1.644~$\mu$m/[{\s}]~1.03~$\mu$m ratio as a function of the
temperature, having fixed the electron density at the derived value of 5
$\times$ 10$^4$ cm$^{-3}$.  The hatched area refers to the HH72A observations,
corrected by $A_\mathrm{V}$=4.7 mag. We assume that all the sulphur is in 
the gas phase and that both species are singly ionized, as suggested by the
absence from the spectra of transitions from any other ionization stage.
Then, the plotted ratio is a function only of the temperature and the
fraction {\it f} of iron in the gas phase.  The observational data agree
with {\it f} values less than 0.4 for all the considered temperatures
(2000 $< T < $ 15\,000 K); in particular, if $T$=10\,000 K, we find 0.20 $<$
{\it f} $<$ 0.35. For comparison, the same analysis applied to HH240A, the
object in the survey with the
richest iron spectrum, gives 0.6 $<$ {\it f} $<$ 1, in substantial
agreement with the abundance estimated in Paper~II by comparing with the
hydrogen recombination lines.\\
Once the iron abundance has been evaluated, it is possible to use the
[{\fe}]~1.644~$\mu$m/[{\n}]~1.04~$\mu$m ratio to estimate the hydrogen
ionization fraction ($x_\mathrm{e}$) and hence the total density of the 
ionized
medium ($n_\mathrm{H}=n_\mathrm{e}/x_\mathrm{e}$). The dependence of the 
above ratio on $x_\mathrm{e}$
assumes that nitrogen is at least partially ionized, as is observationally
testified by the frequent detection in HH objects of the [{\nii}] optical
lines at 6548 and 6583 \AA~(e.g. Bally \& Reipurth 2001). The considered
ratio is plotted in the lower panel of Figure 14 for different values of
the parameter {\it g}, defined as the ratio between the iron abundance and
the fraction of neutral nitrogen.

If we take $T$=10\,000 K, we derive 0.05 $<$ {\it g} $<$ 0.3, which, for
the estimated range of {\it f}, corresponds to $N^0/N$ $>$ 0.67.
Following the procedure described by Bacciotti \& Eisl\"{o}ffel (1999) for
deriving the nitrogen ionization equilibrium, $N^0/N$ $>$ 0.67
corresponds to $x_\mathrm{e}$ $<$ 0.25, and to $n_\mathrm{H}$ $>$ 2.0
$\times$ 10$^5$
cm$^{-3}$. The same procedure applied to HH240A gives 0.85 $<$ 
$x_\mathrm{e}$ $<$
0.95, in agreement with the quite high excitation/ionization conditions
derived for this object.

\subsection{Comparison with predictions of shock models}
In this section, we compare the emission from the observed regions with
the predictions of shock models. As representative cases, we have chosen
the knots richest in lines: HH26A, HH72A and HH320A.  We also model HH25C,
which is the `test case', considered in Section 4.2 above, of an object which
is observed in only $\nu$=1,2 lines.

The main features and the methods used in the shock model computation are
described by Le Bourlot et al. (2002)  and Flower et al. (2003). Here, we
summarize only those aspects most relevant to the present study. The code,
MHD$\_$VODE, simulates one-dimensional, planar, multi-fluid shock waves.
It solves the magnetohydrodynamical equations for the three fluids
(neutral, positively and negatively charged) in parallel with an extensive
chemical network, which links 132 species by 917 reactions.

The manner in which the emission by molecular hydrogen is calculated is
also described by Le Bourlot et al. (2002) and Flower et al. (2003).
Briefly, the processes which populate and depopulate the rovibrational
levels of H$_{2}$ are: collisional excitation and de-excitation;
spontaneous radiative decay; collisional dissociation and ionization; and
reformation on grains, which occurs in the wake of the shock wave. The
equations for the H$_{2}$ level populations are solved in parallel with
the chemical and dynamical conservation equations. This approach is
essential to ensure the accuracy of the computed H$_{2}$ column densities,
because the level populations do not respond instantaneously to changes in
the physical state of the gas. Radiative pumping of H$_2$ was not
taken into account and is considered to be unimportant, compared with
collisional excitation, in the context of the models presented below and
for reasons that will be given there.

In the presence of a transverse magnetic field, a `jump' (J-) type shock
wave will tend to evolve into a `continuous' (C-) type shock wave in
which, prior to the attainment of steady state, a J-type shock remains
embedded (Pineau des For\^{e}ts, Flower \& Chi\`{e}ze, 1997; Smith \& Mac
Low, 1997; Chi\`{e}ze, Pineau des For\^{e}ts \& Flower, 1998). MHD$\_$VODE
can simulate not only steady state J- and C-type shock waves, but also
J-type shocks with a magnetic precursor, as well as the equilibrium
conditions of the preshock gas (see Flower et al. 2003).

In addition to being observable in rovibrational transitions of H$_{2}$,
HH outflows can be traced through the emission of CO and other molecules
and of atoms and ions. In the context of the model, as the temperature and
density of the gas increase, H$_{2}$ is dissociated and H begins to be
ionized. As a consequence, the contribution of radiative cooling by
H$_{2}$ decreases and that of other molecular and atomic species, such as
H$_{2}$O, CO, O, C, C$^{+}$ and Fe$^{+}$, assumes greater importance.
MHD$\_$VODE incorporates the cooling due to rovibrational transitions of
molecules and electronic transitions of atoms, specifically C, N, O, Si,
S, and of their ions, as well as Fe$^{+}$. The strongest ionic/atomic
lines observed are those of [{\fe}] and [{\ci}]; their predicted
intensities are affected strongly by processes within the shock wave. In a
C-type shock wave or a magnetic precursor to a J-type shock, the erosion
of grains can occur owing to the streaming of charged grains through the
neutral gas. At sufficiently high shock speeds, the (refractory) grain
cores can be partially eroded, releasing elements such as Fe and C into
the gas phase: see May et al. (2000).  Neutral iron is ionized rapidly in
the gas phase, predominantly through charge transfer reactions with ions,
such as H$_{3}$$^{+}$ and H$_{3}$O$^{+}$, which have larger ionization
potentials. Fe$^{+}$ may then be excited collisionally and emit the
[{\fe}] forbidden line spectrum.  The predictions of the model with regard
to these two species are considered in Sect. 4.4.6 below, and information
on the atomic data used in the model can be found in the Appendix.

\subsubsection{Model Parameters}
\addtocounter{table}{+3}
\begin{table}[!ht]
\caption{Parameters of shock excitation models}
\begin{tabular}{lcccc}
\hline\\[-5pt]
Object & $v_s$ & $n_\mathrm{H}^{a}$ & Age & Radiation field\\
\hline\\[-5pt]
& km s$^{-1}$ & cm$^{-3}$ & yr & \\
\hline
HH72A  & 31 & 10$^{4}$ & 185 & Yes\\
HH26A  & 52 & 10$^{4}$ & 70  & Yes\\
HH320A & 41 & 10$^{4}$ & 475 & No \\
HH25C$^{1}$ & 45 & 10$^{4}$ &  $>$ 260 & Yes \\
HH25C$^{2}$ & 27 & 10$^{5}$ &  $>$ 680 & Yes \\
\hline\\[-5pt]
\end{tabular}\\
Notes: $^{a}$$n_\mathrm H$ = $n$(H) + 2$n$(H$_{2}$)\\
$^{1}$,$^{2}$Models 1 and 2 of HH25C
\end{table}

It was found that, for HH72A, HH26A and HH320A, neither steady state C-type
nor steady state J-type shocks (which are obtained when the magnetic field
is sufficiently weak) could satisfactorily reproduce the observations. The
former underestimated the column densities of the levels of high
excitation, whereas the latter underestimated the column densities of the
levels of low excitation. However, non-equilibrium models of J-shocks with
magnetic precursors provided good fits to the H$_{2}$ observations of these
three HH objects.  On the other hand, the excitation diagram of HH25C could be
reproduced by a steady state C-type shock model. In this case, the introduction
of a J-component resulted in the populations of levels of high excitation
being overestimated.

The optimal values of the shock speed and preshock gas density (in the
range from 10$^{3}$ to 10$^{5}$ cm$^{-3}$) were determined by comparing
the predicted H$_{2}$ excitation diagram with the observations. The
relation B($\mu$G) = $[n_\mathrm{H}$(cm$^{-3}$)]$^{0.5}$ was adopted as
defining the initial value of the magnetic induction. Two additional
parameters are the age of the shock wave, which determines the extent of
the magnetic precursor, and the strength of the ultraviolet radiation
field.  The model parameters that produce optimal fits to the excitation
diagrams for each object are presented in Table 4.  The entry in the final
column indicates whether the effects of an ultraviolet radiation field
(via photoionization processes) were included in the model. HH72A,
HH26A and HH25C are located in the vicinity of sources of ultraviolet
radiation, and, for these objects, we included
photoionization processes with rates corresponding to the mean
interstellar radiation field in the solar neighbourhood (Draine,
1978). On the other hand, there appear to be no sources of ultraviolet
radiation nearby HH320A, and, in this case,
photoionization processes were neglected. It is perhaps
significant that the inclusion of a radiation field improved the fits to 
the empirical excitation diagrams for HH72A, HH26A and HH25C but made worse 
the fit for HH320A.

The rate of photodissociation of H$_2$ by the mean interstellar
radiation field is of the order of 10$^{-11}$ s$^{-1}$ (cf. Abgrall et al.
1992).  The same process (of photon absorption) which gives rise to
photodissociation also gives rise to H$_2$ fluorescence through radiative
cascade from the excited (optically pumped) electronic states to the
rovibrational levels of the electronic ground state.  Optical pumping
followed by fluorescence is an order of magnitude more probable than
optical pumping followed by dissociation of H$_2$.  Thus, the total rate
of the fluorescence cascade is of the order of 10$^{-10}$ s$^{-1}$.  On
the other hand, any given rovibrational level of the electronic ground
state receives only a fraction of the cascades.  We take 10$^{-11}$
s$^{-1}$ as an order of magnitude estimate of the rate of population of
any given excited rovibrational state through the optical pumping by the
unshielded mean interstellar radiation field.  We shall now compare this
rate with that of collisional excitation. 

At kinetic temperatures of a few thousand kelvin, attained in the shock
wave, the rate coefficients for rovibrational excitation of H$_2$ by H are
typically of the order of 10$^{-12}$ cm$^{-3}$ s$^{-1}$ and exceed those
for excitation by He and H$_2$ by more than a factor of 10 (cf. Le Bourlot
et al. 1999). Furthermore, the fractional abundance of H in the shock wave
becomes comparable with that of He, i.e. of the order of 0.1.  In the
models considered below, the preshock gas density $n_\mathrm{H} \geq$
10$^4$ cm$^{-3}$.  Thus, the rate of rovibrational excitation of H$_2$ by
H is of the order of 10$^{-9}$ s$^{-1}$ or greater.  We conclude that the
inclusion of optical pumping would not have a significant effect on
computed column densities of the excited rovibrational levels of H$_2$ and
hence on their comparison with the observations.

Low preshock gas densities result in low rates of collisional excitation
of H$_{2}$ molecules, smaller excited state column densities and greater
departures of the level populations from LTE. For a given density, the
shock speed is the main factor determining the column densities: an
increase in the shock speed results in larger column densities. Indeed,
the model is so sensitive to the shock speed that a change of the order of
1 km s$^{-1}$ in $v_s$ can make a significant difference to the excitation
diagram. The age of the outflow affects differentially the population
densities of the more highly excited levels: the younger the object, the
closer is the shock wave to being J-type and the higher are the column
densities of the more highly excited rovibrational levels. Furthermore,
the populations of these levels tend to be closer to LTE.  The presence of
a radiation field leads to an increase in the degree of ionization of the
gas and hence in the strength of the coupling between the neutral and the
charged fluids; there results a narrower shock wave.

The observations comprise emission lines from H$_2$ levels of both
ortho (J odd) and para (J even) symmetry and hence reflect the variation
of the ortho:para ratio through the shock wave.  In the hot gas produced
by the shock wave, para-H$_2$ is converted to ortho-H$_2$ by nuclear spin
changing reactions with H.  As shown by Wilgenbus et al. (2000), the local
ortho:para ratio increases from a low value in the cold preshock gas
(assuming that it has become thermalized through proton transfer reactions
with H$^+$ and H$_3^+$) to a value of 3 immediately behind the shock wave.
The ratio of the total column densities in the ortho and para levels,
N(ortho)/N(para), lags behind the local value of the ratio of the number
densities of the ortho and para levels.  However, our observations do not
yield N(ortho)/N(para), as the contributions from the levels of the
vibrational ground state, whose populations become significant in the
cooling flow, are not included.  Rather, the observations (of excited
vibrational levels) reflect the value of the ortho:para ratio in the
region of the hot gas in which the corresponding levels are preferentially
excited. Thus, we consider that any attempt to derive a unique `ortho:para
ratio' from our observations would be misguided.  On the other hand, any
systematic departure of the ortho:para ratio from its value in LTE would
manifest itself as a downwards displacement of the ortho levels in the
excitation diagram (cf. Wilgenbus et al. 2000); no such shift is
discernible in the comparison with the models presented below.

\subsubsection{HH72A}
\begin{figure}[ht]
\includegraphics[width=8cm]{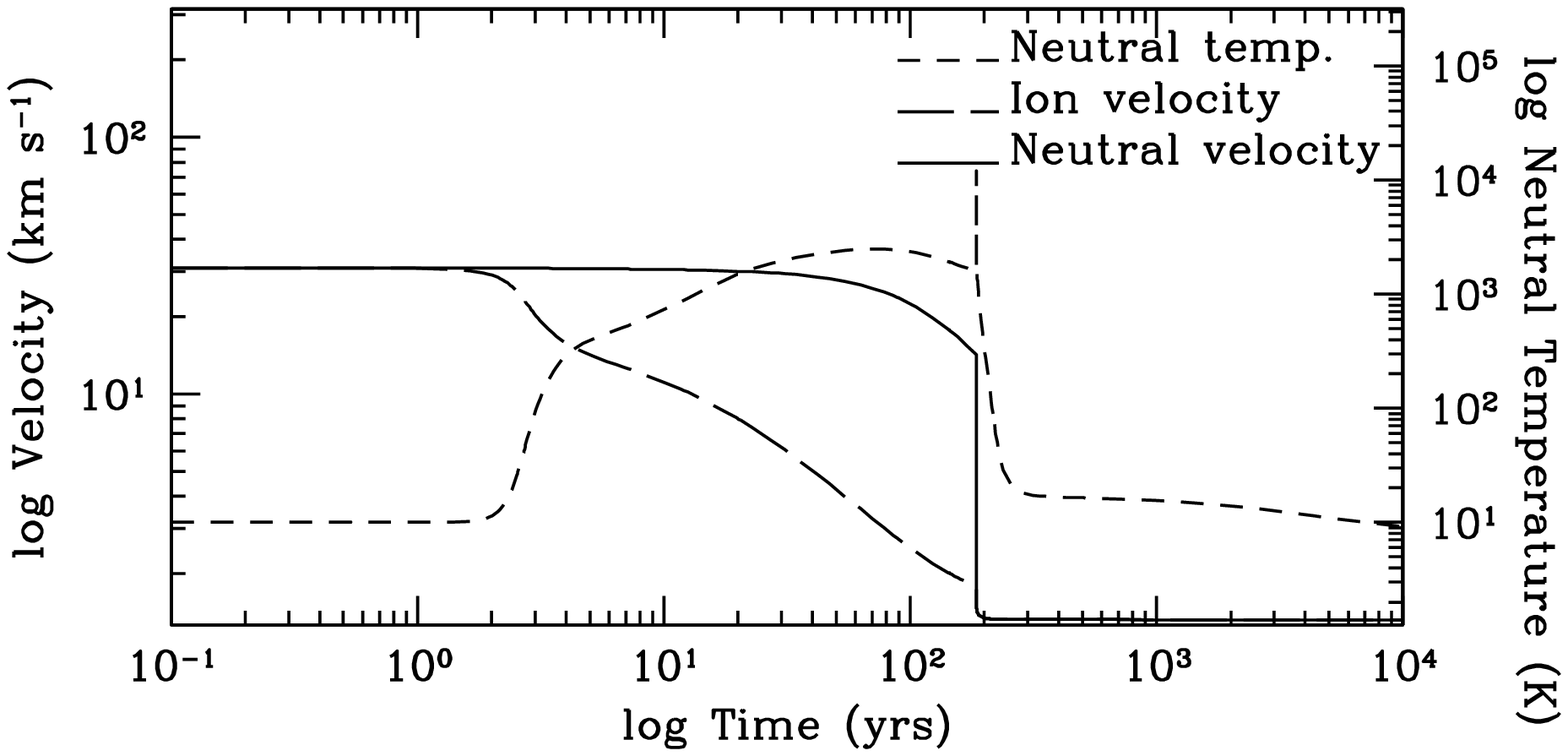}
\caption[]{Velocity and temperature profiles of the model of HH72A. The
velocity is expressed in the shock frame
and the preshock gas is to the left. The J-shock discontinuity is evident
in both the temperature and velocity profiles.
\label{72Aprof}}
\end{figure}
\begin{figure*}[ht]
\centering
\includegraphics[width=12cm]{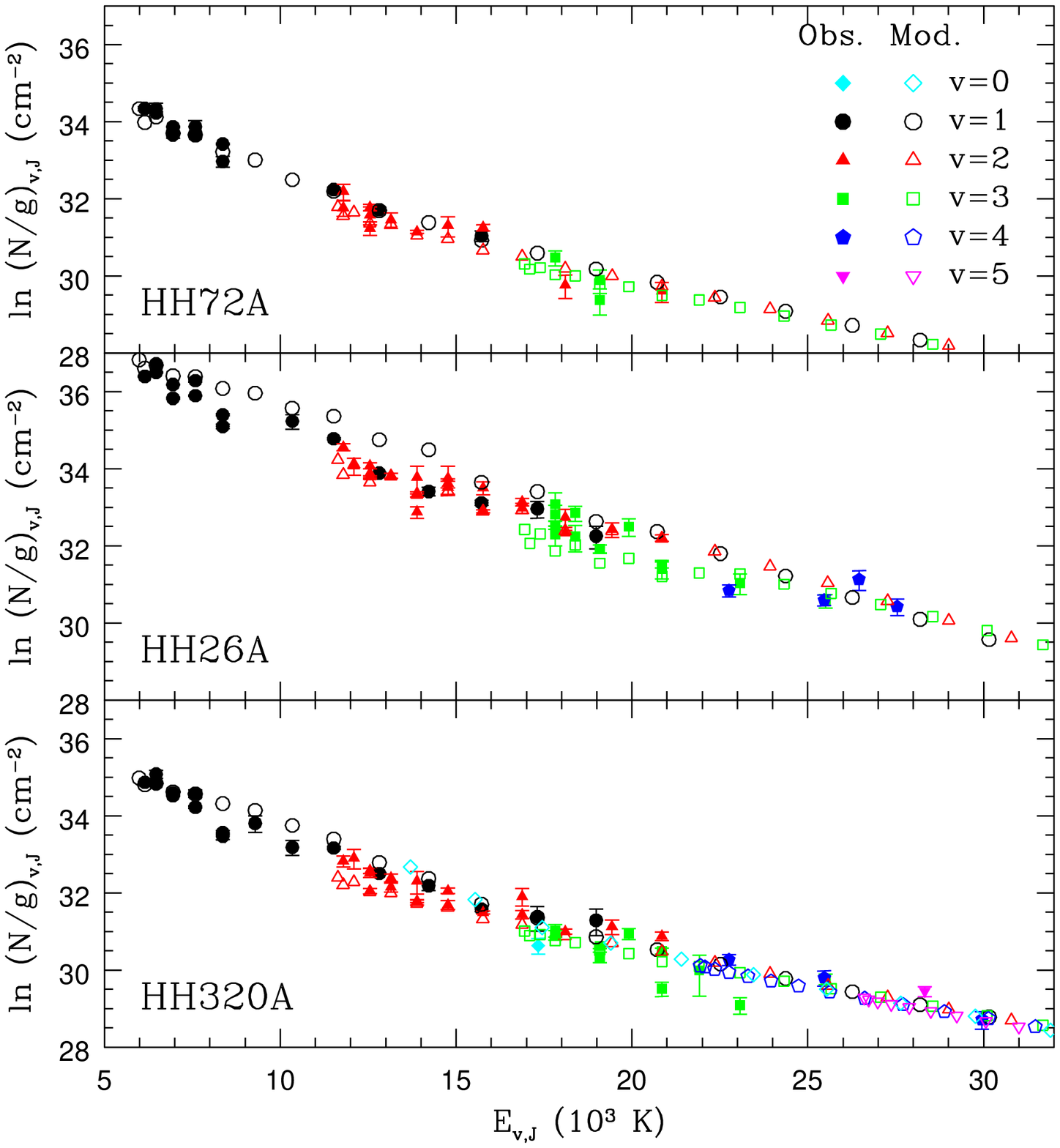}
\caption[]{Models predictions for the excitation diagrams of HH72A, HH26A
and HH320A.
Different symbols distinguish transitions coming from different manifolds.}
\end{figure*}

The velocity and temperature profiles of the model for HH72A are
plotted against flow time, $t_n = \int \frac{1}{v_n} dz_n$, in Figure 15:  
the time at which the J-type discontinuity occurs may be identified with
the age of the shock wave. The models of HH26A and HH320A exhibit
qualitatively similar velocity and temperature profiles, which will not be
shown.

In Figure 16 (top panel) we show the excitation diagrams derived from the
model (open symbols) and the observations (corresponding filled symbols).
The observations extend up to approximately 2 $\times$ 10$^{4}$ K above
ground and include $\nu$=1, 2 and 3 vibrational levels.
At successive vibrational thresholds, the observational points do not
depart notably from the median curve, indicating that there is a high
degree of thermalization at the local temperature. Models with initial
densities less than 10$^{4}$ cm$^{-3}$ yield lower column densities which
diverge more evidently from the median curve at each vibrational
threshold. Conversely, models with initial densities higher than 10$^{4}$
cm$^{-3}$ produce level populations which are closer to LTE, but the
curvature of the Boltzmann plot is not consistent with that observed.
Similar conclusions were reached from the analysis of the spectra of both
HH26A and HH320A.

The characteristic length of the shock wave is determined by the distance
over which the ion and neutral velocities differ. For HH72A, this length
is approximately 2 $\times$ 10$^{16}$ cm, which may be compared with the
observed dimension of the source, 3.5 $\times$ 10$^{16}$ cm (at a distance
of 1500 pc and for an angular size of 1.5 arcsec, as measured on the image
taken at the telescope). The same comparisons for HH26A and HH320A
produced similar levels of agreement.

\subsubsection{HH26A}

The observed points in the H$_{2}$ excitation diagram extend, in this
case, to almost 3 $\times$ 10$^{4}$ K and include levels of the first four
excited vibrational states: see Figure 16, middle panel. The column
densities are larger than observed in HH72A, implying a higher shock
speed, and show a similar degree of departure from LTE.  The higher shock
speed used for HH26A results in a higher rate of H$_{2}$ cooling and so
the shock is narrower than HH72A; it is for this reason that the model of
HH26A has a smaller age than HH72A.

The model of HH26A yields column densities of levels with $\nu$ = 1 which
exceed the observations beyond approximately 10$^{4}$ K (cf. Figure 16);
this is symptomatic of a more general tendency of the models to
overestimate the populations of the levels in the $\nu$ = 1 manifold. Such
a tendency may point to an overestimation of the rates of vibrationally
inelastic $\nu$ = 0 $\rightarrow$ $\nu$ = 1 collisions.  The model of
HH26A has the highest shock speed of the three objects that we consider
and hence has the largest fractional abundance of H, owing to collisional
dissociation of H$_{2}$.  As the rate coefficients for vibrational
excitation of H$_{2}$ by H are much larger than for excitation by He or
H$_{2}$, a larger H abundance exacerbates the effect of any systematic
error in the H-H$_{2}$ collisional rate coefficients. However, we stress
that there is no independent evidence to confirm the existence of such a
systematic error.

\subsubsection{HH320A}

Levels of the first six vibrational states, with excitation energies
extending to over 3 $\times$ 10$^{4}$ K, are observed: see Figure 16,
bottom panel. The column densities observed for two levels with $\nu$ = 3
($E$ = 20\,857 K and $E$ = 23\,070 K) fall significantly below the mean
curve in the excitation diagram.  The level at $E$ = 20\,857 K has been
observed twice, in two different transitions: one observation falls on the
median curve, while the other, as already stated, falls below it and out
of line with the remaining $\nu$ = 3 levels.  We believe that this
discrepancy arises from systematic errors (related to atmospheric
absorption) in the observations and that the misalignment of the
observational point at $E$ = 23\,070 K level is attributable to a similar
effect. The absence of a radiation field in this model results in a wider
shock wave than is the case for the other objects and hence to a greater
shock age.

No atomic or ionic emission is observed towards the BHR71 region and
the absence of both [{\fe}] and [{\ci}] lines, which were observed towards
all other outflows considered in this paper and in papers I and II, is
particularly surprising; here we discuss why these lines are not observed.

Perhaps the most extreme possibility is that there is very little Fe and C
present in the gas phase of the BHR71 region because they are mainly in
solid (and, in the case of C, in molecular) form, which would imply that
the dust grains have undergone little erosion.  Whilst such a situation
cannot be discounted by our observations, we consider that it is unlikely
because, according to our modelling results, the gas has been
subjected to a strong shock.  Emission from [{\ci}] and [{\fe}] would not
be observed if C and Fe$^+$ were ionized by collisions, photons or charge
transfer reactions; but it should be noted that we observe no lines of
Fe$^{2+}$ in this region.  The energies required for ionization
of C and Fe$^{+}$ (11.26 eV and 16.16 eV, respectively) are such that, if
collisional or photoionization is invoked, we should expect emission from
other ions (for example, [S II] and [N II]) that are not observed.  
Charge transfer ionization of C to C$^+$ is ineffective and Fe$^{2+}$
rapidily recombines via charge transfer recombination reactions with H
(Kingdon and Ferland, 1996).  For these reasons, we consider that
ionization is unlikely to be responsible for the absence of [{\ci}] and
[{\fe}] emission.  Collisional deexcitation of the lines could be
responsible for their relative weakness.  The rate of electron collisional
deexcitation of the [{\fe}] and [{\ci}] lines is of order 10$^{-7}$ and
10$^{-9}$ cm$^3$ s$^{-1}$, respectively: by comparison, the radiative
decay rates are of order 10$^{-3}$ s$^{-1}$ for [{\fe}] and 10$^{-4}$
s$^{-1}$ for [{\ci}].  Therefore, in order for collisional deexcitation to
dominate radiative decay, $n_\mathrm{e}$ must be greater than 10$^4$ or
10$^5$ cm$^{-3}$, respectively, which would imply that hydrogen is fully
ionized.  An alternative and more probable explanation is that the atomic
and ionic emission arises in the apex of a bow shock or a reverse shock in
the jet.  In either case, the region of atomic and ionic emission would be
small, and it is possible that the slits did not encompass fully the
emitting areas.

\subsubsection{HH25C}

As noted in Sect. 4.2, HH25C is an object which is observed in $\nu$=1,2
lines, at excitation energies extending up to $\approx$ 1.5 $\times$
10$^{4}$ K, beyond which we have only upper limits on the line intensities
and the column densities of the corresponding levels.

\begin{figure*}[ht]
\centering
\includegraphics[width=12cm]{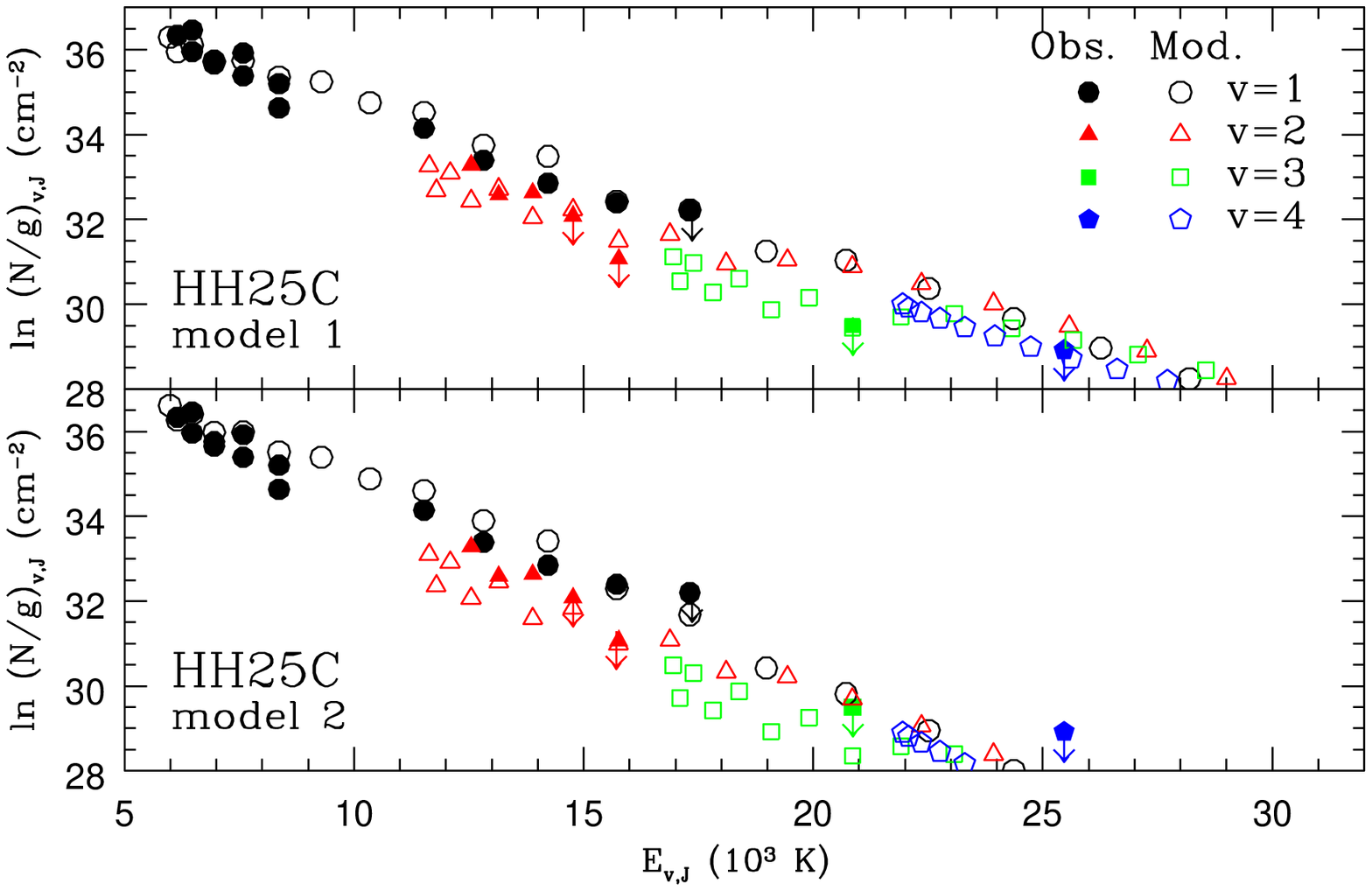}
\caption[]{Excitation diagram of HH25C: the observational data are compared
with predictions from two different steady-state C-shock models (1 and 2)
whose parameters are listed in Table 4. The arrows indicate 3$\sigma$ upper
limits on the column densities of the corresponding transitions.}
\end{figure*}

We present two models of HH25C, whose parameters are given in Table 4.
Both the models are of C-type shocks in steady state, and the flow time
through the shock wave provides only a lower limit to its age. The models
are compared with the observations of HH25C in Figure 17. Regarding the
levels which are definitely observed, with excitation energies less than
about 1.5 $\times$ 10$^{4}$ K, the model with the lower density (and
higher shock speed) provides perhaps a better fit: the column densities of
the $\nu$=1 levels are overestimated and those of the $\nu$=2 levels
are underestimated to a greater extent by the higher density (lower
velocity) model 2. On the other hand, the predictions of the higher
density model are more consistent with the observational upper limits at
excitation energies above 1.5 $\times$ 10$^{4}$ K. Straight line fits to
the theoretical data yield excitation temperatures of 2770 K for model 1
and 2130 K for model 2; the latter result is more consistent with the
value of $\approx$ 2300 K obtained by fitting to the observational data.

\subsubsection{[{\fe}] and [{\ci}] emission lines}

Emission lines of [{\fe}] and [{\ci}] were also observed towards HH72A and
HH26A. For both objects, the intensities predicted by the models (which
successfully reproduce the H$_{2}$ excitation diagrams) are more than 2
orders of magnitude (in the case of [{\fe}]) and 5 orders of magnitude (in
the case of [{\ci}]) smaller than are observed.  

An important difference between the shock wave model and the NLTE model
used in Sect. 4.3 is that the shock model predicts a low ionization
fraction. For example, the ionization fraction within the shock wave model
of HH72A does not exceed about 2 $\times$ 10$^{-5}$.  This degree of
ionization is insufficient for electron collisions to play a significant
role. The excitation of Fe$^{+}$ and of C occurs principally in collisions
with neutral perturbers, which are less effective than those with
electrons.  For the rates of collisional excitation of [{\fe}] by
electrons to be comparable with the rate of excitation by neutrals, the
ratio $x_\mathrm{e} = n_\mathrm{e}/n_\mathrm{H} \approx 10^{-3}$. In
order to enhance the computed [{\fe}] line intensities to values
comparable with those observed requires $x_\mathrm{e} \approx$ 0.1;  
this may be compared with the upper limit $x_\mathrm{e}$ $<$ 0.25 derived
in Sect. 4.3.

The fact that the models underestimate the [{\fe}] and [{\ci}] line
intensities suggests that these forbidden lines are emitted from regions
where the degrees of dissociation and ionization are much higher, which do
not contribute significantly to the H$_{2}$ line intensities. Such regions
could be the apex of a bow shock (with the H$_{2}$ emission being produced
in the wings), a reverse shock in the jet, or photoionized gas. However,
the last of these three possibilities would imply local sources of
ultraviolet radiation, sufficiently hard and intense to photoionize
hydrogen; we know of no observational evidence for the existence of such
sources of radiation.  Accordingly, we lean towards an explanation in
terms of the collisional ionization of hydrogen, in high velocity J-type
shocks. Indirect evidence that the [{\fe}] and H$_{2}$ emission lines do,
indeed, arise in distinct regions is afforded by the difference in the
values of the extinction which are derived from the respective line
intensities (at least in the case of HH72A) and from the high values of
the electron density derived from the [{\fe}] lines.
In future studies, we intend to address the issue of the origin the atomic
and ionic forbidden lines in a more quantitative manner, within the
framework of a shock model.

\section{Conclusions}
We have presented the 0.95-2.5~$\mu$m spectra of three protostellar jets
(HH24-26, HH72, HH320-321).
The analysis of the observed features has led to the following conclusions:
\begin{itemize}
\item[-]
Strong H$_2$ emission lines with excitation energies up to 35\,000 K are
detected throughout all the Herbig-Haro objects present in the three
regions, while infrared knots show lines with excitation energies rarely
exceeding 15\,000 K. This difference in the line emission reflects
different excitation regimes: while the condensations observed only in the
infrared are excited at a single temperature of $\approx$\,2000-3000 K,
Herbig-Haro objects have a temperature stratification, with components up
to more than 5000 K, which tend towards the values (of the order of
10$^4$ K) determined from optical transitions. We note that, in order to
trace effectively the highest temperature components, it is essential to
obtain observations in the 1.0-1.4 $\mu$m range, where the lines with high
vibrational quantum numbers are located.

\item[-]
Atomic and ionic emission in form of [{\fe}], [{\s}] and
{[{\n}]} transitions is detected in only a few HH objects. The ratios
of lines of different species have been used to derive some of the
physical parameters of the ionized gas (electron density, iron abundance
and ionization fraction), which all indicate that the shocks present in
these objects have lower excitation/ionization conditions than in cases
where stronger [{\fe}] emission occurs.

\item[-]
The observed H$_2$ emission in the HH objects can be reproduced by
models of J-type shocks
with magnetic precursors and ages of typically a few hundred years. The
shock speeds are in the range from about 30 to about 50 km s$^{-1}$,
and the preshock gas density is of the order of 10$^4$ cm$^{-3}$.
 In the case of the pure H$_2$ knot HH25C, it is possible to fit marginally the
observational data by means of a steady state (C-type) shock model
with a higher preshock density (10$^5$ cm$^{-3}$).

\item[-]
The same planar model that fits the H$_{2}$ emission systematically
underestimates both the [{\fe}] and [{\ci}] lines fluxes observed in HH72A
and HH26A. We believe that either a bow shock (with atomic and ionic
emission originating at the apex of the shock and with H$_{2}$ arising in
the wings) or a reverse shock in denser gas at the centre of the outflow
is responsible for the ionic and atomic emission. Supporting evidence
comes from the high electron density ($\approx$ 5 $\times$ 10$^4$
cm$^{-3}$) and visual extinction derived from the [{\fe}] lines.

\end{itemize}

\emph{Acknowlegements}: We thank an anonymous referee for a detailed
and constructive report.

\section{Appendix: Fe$^{+}$ and C$^{0}$ models.}
In the model of Fe$^{+}$, we considered transitions among the 19 energy
levels which arise from the a$^{6}$D, a$^{4}$F, a$^{4}$D, a$^{4}$P
and b$^{4}$P terms. All of the observed infrared transitions are
emitted from a$^{4}$D; the strongest lines are at 1.644 $\mu$m
(a$^{4}$D$_{\frac{7}{2}}$ - a$^{4}$F$_{\frac{9}{2}}$) and
1.257 $\mu$m (a$^{4}$D$_{\frac{7}{2}}$ - a$^{6}$D$_{\frac{9}{2}}$).
In order to ensure that the populations of
the levels of the a$^{4}$D term, which lies about 1 eV ($\approx$ 10$^{4}$
K) above ground, are converged, we included cascades from the
a$^{4}$P and b$^{4}$P terms; the b$^{4}$P triplet
lies approximately 3 eV ($\approx$ 3 $\times$ 10$^{4}$ K) above ground.

The rates of collisional and radiative transitions determined the Fe$^{+}$
level populations at each point in the model. The spontaneous radiative
transition probabilities computed by Quinet, Le Dourneuf \& Zeippen (1996)
were adopted. Collisions with both neutral species and
electrons were included; for the latter, we used the collision strengths of
Zhang \& Pradhan (1995).  Regarding the neutral perturbers, we included
collisions with the most abundant species, H$_{2}$, H and He.
However, we are not aware of any quantum mechanical calculations of
cross sections for the electronic excitation of Fe$^{+}$ by these
perturbers.  Accordingly, we estimated the rate coefficients using a simple
classical approximation, which derives from the Langevin `orbiting' model,
adopting the  polarizability and reduced mass appropriate to each perturber.

Assuming LS coupling (cf. Nussbaumer \& Storey 1980), the conservation
of the total electron spin implies that the spin state of Fe$^{+}$ can be
changed only by exchange of a bound and an incident electron with
oppositely directed spins [if (weak) magnetic interactions are neglected].
Electron collisional excitation of the a$^{4}$D term from the a$^{6}$D
ground term proceeds via this electron exchange mechanism. In collisions
between Fe$^{+}$ and H, a similar process of exchange can, in principle,
take place, between one of the bound electrons of the ion and the bound
electron of the hydrogen atom. In the case of collisions with He, on the
other hand, the ground state is a singlet, and exchange with an electron
of opposite spin implies a transition to a triplet state. As the lowest
energy triplet state of He lies almost 20 eV above ground, such events
will be extremely rare under the physical conditions that we are
considering. Finally, ground state H$_{2}$ is also a singlet, and the
lowest energy triplet state is repulsive. Hence, transitions to the
triplet state lead to dissociation of the molecule and require an energy
input of more than 4.5 eV. We conclude that transitions involving a change
of spin, including excitation of a$^{4}$D from a$^{6}$D, are induced at a
significant rate only by collisions with electrons and H atoms.

Collisional and radiative transitions among the five energy levels of
neutral C which arise from the $^{3}$P, $^{1}$D and $^{1}$S
terms were incorporated in the model.  Following the same reasoning as
above, we should include all levels of C$^{0}$ which contribute
significantly, by direct collisional excitation or cascade, to the rates
of population of the levels from which transitions are observed.  The
uppermost $^{1}$S term has an energy of 3.1 $\times$ 10$^{4}$ K,
relative to ground. The 0.983 $\mu$m and 0.985 $\mu$m lines originate in
the $^{1}$D term, which has an energy of 1.6 $\times$ 10$^{4}$ K.
Population transfer in collisions with H (Launay \& Roueff 1977), H$_{2}$
(Schr\"{o}der et al. 1991), He (Staemmler \& Flower 1991), and H$^{+}$
(Roueff \& Le Bourlot 1990) was taken into account. The data pertaining to
electron collisions and also the radiative transition probabilities were
taken from the compilation of Mendoza (1983).

\end{document}